\documentclass[a4paper,fleqn,usenatbib]{mnras}

\usepackage{newtxtext,newtxmath}

\usepackage[T1]{fontenc}
\usepackage{ae,aecompl}

\usepackage{graphicx}	
\usepackage{amsmath}	
\usepackage[dvipsnames]{xcolor}

\newcommand{\plong}{P_{\rm L}}	
\newcommand{\pshort}{P_{\rm S}}
\newcommand{\pslr}{P_{\rm S}/P_{\rm L}}	
\newcommand{\pov}{P_{\rm 1O}}	
\newcommand{\poov}{P_{\rm 2O}}	
\newcommand{\pooov}{P_{\rm 3O}}	
\newcommand{\pf}{P_{\rm F}}	
\newcommand{\px}{P_{\rm x}}	
\newcommand{\psh}{P_{\rm sh}}	
\newcommand{\pmodul}{P_{\rm m}}	
\newcommand{\aov}{A_{\rm 1O}}	
\newcommand{\aoov}{A_{\rm 2O}}	
\newcommand{\aooov}{A_{\rm 3O}}	
\newcommand{\af}{A_{\rm F}}	
\newcommand{\ax}{A_{\rm x}}	
\newcommand{\ash}{A_{\rm sh}}	
\newcommand{\fov}{\nu_{\rm 1O}}	
\newcommand{\foov}{\nu_{\rm 2O}}	
	
\newcommand{\ff}{\nu_{\rm F}}	
\newcommand{\fx}{\nu_{\rm x}}	
\newcommand{\fsh}{\nu_{\rm sh}}	
\newcommand{\fmodul}{\nu_{\rm m}}	
\newcommand{\sn}{{\mathrm{S/N}}}
\newcommand{\cd}{{\,\mathrm{c\,d^{-1}}}}	
\newcommand{\idlmc}[1]{LMC-#1}	
\newcommand{\idsmc}[1]{SMC-#1}

\title[1O Cepheids -- beyond radial modes]{First Overtone Cepheids of the OGLE Magellanic Cloud Collection -- beyond radial modes}

\author[Smolec et al.]{
R. Smolec,$^{1}$\thanks{E-mail: smolec@camk.edu.pl}
O. Zi\'o\l{}kowska,$^{1,2}$
M. Ochalik,$^{3}$
and M. \'Sniegowska$^{1,4}$
\\
$^{1}$Nicolaus Copernicus Astronomical Center, Polish Academy of Sciences, Bartycka 18, 00--716 Warsaw, Poland\\
$^{2}$Astronomical Observatory, University of Warsaw, Al. Ujazdowskie 4, 00-478 Warszawa, Poland\\
$^{3}$Astronomical Observatory, Jagiellonian University, Orla 171, 30-244 Krak\'ow, Poland\\
$^{4}$School of Physics and Astronomy, Tel Aviv University, Tel Aviv 69978, Israel
}

\date{Accepted XXX. Received YYY; in original form ZZZ}

\pubyear{2020}

\begin{document}
\label{firstpage}
\pagerange{\pageref{firstpage}--\pageref{lastpage}}
\maketitle

\begin{abstract}
We have analysed Optical Gravitational Lensing Experiment photometry for first overtone classical Cepheids in the Large and Small Magellanic Clouds in search for additional periodicities beyond radial modes. We have used standard consecutive prewhitening technique in some cases followed by time-dependent prewhitening. We report new candidates for double-mode radial pulsations. However, majority of signals we have detected cannot be interpreted in terms of radial modes. We report 516 double-periodic stars with period ratios, $\px/\pov$, in the range 0.60 and 0.65. We study the properties of this class and implications for model explaining these periodicities. We also report 28 stars in which additional variability is of longer period, below radial fundamental mode, with median period ratio, $\pov/\px$, of 0.684. This class is an analogue of a class known in RR Lyrae stars. Hundreds of other signals were detected that cannot be attributed to radial modes or the above-mentioned classes. Statistical properties of these signals are analysed. We suggest that majority of these signals correspond to non-radial modes. In particular, a significant fraction can be attributed to non-radial modes of moderate degrees, tightly connected to a class with period ratios in between 0.60 and 0.65. In tens of stars, close to radial mode frequency, relatively large-amplitude and coherent signals are observed, that may represent yet another class. In 27 stars periodic modulation of pulsation was detected. Differences in additional frequency content between the two Clouds are discussed.
\end{abstract}



\section{Introduction}

Classical Cepheids (hereafter Cepheids) are among the most important tools of stellar astrophysics. Thanks to the period-luminosity relation, tight especially in the near-infrared bands \citep[see eg.,][]{BreuvalPL,RipepiPL,OwensPL}, they are among the most important distance indicators, essential for measuring the Universe and determining its expansion rate \citep[eg.][]{Freedman2012,Riess2019}. Classical Cepheids pulsate either in radial fundamental (F) mode or in the radial first overtone (1O) mode or in the radial second overtone (2O) mode. Double-mode pulsators, involving different pulsation modes (F+1O, 1O+2O, 1O+3O) or even triple-mode pulsators are also known and proved to be useful tools to test the stellar pulsation and evolution theories, or to constrain physical parameters of these important variables \citep[eg.][]{md05,Beaulieu2006,Pilecki2018,DeSomma2020}. 

While single-mode large-amplitude radial pulsation of Cepheids is well understood, the mechanisms behind double-mode and triple-mode radial pulsation remain unclear \citep[see][for opposing views]{kollath2002,sm08b}. In the recent years, classical Cepheids revealed even more complex nature through the presence of plethora of additional low-amplitude periodicities that cannot be interpreted as due to radial pulsation. These low-amplitude periodicities appear to be common in first overtone pulsators and scarce in the fundamental mode Cepheids. 

The discoveries of low-amplitude periodicities in classical Cepheids were possible thanks to long-term continuous monitoring of these stars by the Optical Gravitational Lensing Experiment \citep[OGLE;][]{o3,o4}. The OGLE Collection of variable stars (OGLE-CVS) includes virtually all classical Cepheids in the Magellanic Clouds \citep{o4_clouds_henrietta} and for majority of these stars high quality, homogeneous data, covering several observing seasons, is available, allowing us to detect periodic signals at the milli-magnitude level.

 \cite{mkm04} and \cite{mk09} analysed OGLE LMC data and reported that in 8\,per cent of 1O Cepheids additional low-amplitude signals with frequencies close to the radial mode frequency are detected and concluded that these signals are most likely due to excitation of non-radial modes. Additional signals are coherent. Sometimes two peaks are detected in the frequency spectrum on the same side of the radial mode frequency. 

The most interesting new form of double-periodic pulsation are classical Cepheids with dominant 1O variability and with additional low-amplitude variability of shorter period, with period ratios $\px/\pov$ in the $(0.60,\,0.65)$ range (conversely, with frequencies in the $(1.54,\,1.67)\fov$ range). The group was first identified through analysis of LMC data by \cite{mk08,mk09} and \cite{o3_lmc_cep}, with two sequences of slightly different period ratios reported in the latter study. In the SMC, \cite{o3_smc_cep} reported 138 double-periodic stars forming three distinct sequences in the Petersen diagram. This sample was analysed in detail by \cite{ss16}, who found additional signals in 35\,per cent of these stars, centered at sub-harmonic frequencies, $1/2\fx$. Interestingly, double-periodic pulsation of the same characteristics exists among first overtone RR~Lyr (RRc) stars, see eg., \cite{ns19} and was recently detected in an anomalous Cepheid with {\it TESS} \citep{PlachyTESS}. A common explanation was proposed by \cite{wd16} who associated the signals at $\fx$ with harmonics of non-radial modes of moderate degrees. The three sequences in the Petersen diagram for Cepheids were associated with excitation of non-radial modes of $\ell=7$ (top sequence), $\ell=8$ (middle sequence) and $\ell=9$ (bottom sequence). In this model, direct detection of non-radial modes, at $1/2\fx$ is less likely, as expected amplitude is low due to geometric cancellation. The expected amplitude is the largest for even $\ell$ and indeed majority of the detections at $1/2\fx$ correspond to the middle sequence in the Petersen diagram \citep{ss16}. We also note that beyond Magellanic Clouds, this form of pulsation was also detected in OGLE Galactic disk and Galactic bulge data \citep{Pietruk2013,Rathour2021} and recently in {\it TESS} data \citep{PlachyTESS}.

As the just described form of double-periodic pulsation is present both for RRc stars and 1O Cepheids, one may speculate that other forms of double-periodic pulsation discovered for RRc stars may also be present among 1O Cepheids. \cite{nsd15} reported the discovery of a puzzling group of double-periodic RRc stars, with additional periodicity of longer period; in the Petersen diagram, these double-periodic stars cluster tightly at $\pov/\px=0.686$. Additional periodicity is always coherent in these stars. The nature of this group remains puzzling \citep[see][for discussion]{wd16}. Double-periodic 1O Cepheids that share similar characteristic were reported by \cite{sa18b} who analysed the whole OGLE LMC and SMC sample in search of additional periodicities.

Other form of variability detected in classical Cepheids is periodic modulation of pulsation. \cite{mkm04} discovered long-period modulation of double-mode 1O+2O Cepheids. As analysed by \cite{mk09}, both modes are modulated with the same period and their amplitudes are anti-correlated. Recently modulation was detected in a double-mode, F+1O Cepheid \citep{Rathour2021}. Modulation was also detected in single-mode Cepheids: in the 2O Cepheid V473~Lyr \citep[see][]{MolnarSzabados}, in a few 1O Cepheids \citep{o4_clouds_multimode, Kotysz, Rathour2021} and in several single-mode F-mode Cepheids in the LMC \citep{s17} and one Cepheid in the Galactic disk \citep{Rathour2021}. Interestingly, for single-mode F-mode Cepheids, except these modulations, additional periodicities are hardly present in the frequency spectra, in sharp contrast to 1O pulsators \citep{s17,sa18b}.

In this paper, we conduct in-depth frequency analysis of 1O Cepheids from the OGLE-IV LMC and SMC sample using classical consecutive prewhitening technique, supplemented with the time-dependent prewhitening for non-stationary variations. For each star with additional variability beyond radial mode, manual analysis is conducted to extract all low-amplitude periodicities present in the data. Similar analysis was done by \cite{sa18b}, who used kernel-regression method for prewhitening \citep{sa18a} but searched only for one single additional periodicity.

Some preliminary results of this study were presented in \cite{MSniegProc} and \cite{OZProc}.


\section{Data analysis}

We analyse OGLE-IV \citep{o4} $I$-band data for first overtone classical Cepheids in the SMC and LMC included in the OGLE Collection of Variable Stars \citep[OGLE-CVS,][]{o4_clouds}. The sample consists of 1784 stars in the SMC and 1766 stars in the LMC\footnote{The collection is regularly updated with new stars. Current numbers may be slightly higher.}. We use a standard consecutive prewhitening technique \citep[see eg.,][]{mk09} to search for additional periodicities beyond the dominant radial first overtone mode. Significant periodicities are identified with the help of the discrete Fourier transform (DFT) and are included in a sine series of the following form 
\begin{equation}
m(t)=A_0+\sum_{k} A_k\sin(2\pi \nu_k t+\phi_k)\,,\label{eq:sseries}
\end{equation}
fitted to the photometric data, $m(t)$. Above, $A_0$ is mean magnitude, index $k$ enumerates the sine terms for which $A_k$ and $\phi_k$ are amplitude and phase corresponding to variability with frequency $\nu_k$. The dominant terms correspond to Fourier series describing radial first overtone variability ($\fov$ and its harmonics). Residuals from the fit are then inspected with DFT to identify further periodicities of lower amplitude. In principle, signals with $\sn>4$ were considered significant and included in the solution. Weaker signals, with $\sn>3.5$, were also included in the solution, provided these were linear combinations of the already identified frequencies or were located at the expected frequency (eg. close to the sub-harmonic frequencies). 4-$\sigma$ outliers were successively removed from the data. Slow trends were modelled either using low-order polynomials or spline functions. In some cases time-dependent prewhitening (see below) was employed to remove slow trends. Two frequencies were considered resolved, provided their difference exceeds $2/\Delta T$, where $\Delta T$ is data span.

First, we applied an automatic procedure to identify stars which only show 1O pulsation and no signature of additional periodicity. To this aim, Fourier series was fitted to the data to the order satisfying $A_k/\sigma(A_k)>4$. Possible slow trends were modelled with second order polynomial function and severe (6-$\sigma$) outliers were removed from the data. DFT of the residual data was searched for significant signals in the $(0,\,6\fov)$ range. A star without significant signals was not analysed further. All other stars were subject to detailed manual analysis.

When new significant signal is detected in the data, we first check whether it can be represented as a linear combination of the previously detected frequencies. If so, no new independent frequency is introduced, just respective combination term is included in the solution. For low-amplitude double-periodic pulsation the most common combination is $\fx+\fov$. A special case is the periodic modulation of pulsation, which, in general, manifests as equidistant multiplets centred on radial mode frequencies, and on the harmonics, as well as through signal at the modulation frequency (and its harmonics), see eg., \cite{bsp11}. Separation within multiplet components corresponds to modulation frequency, $\fmodul=1/\pmodul$. In the ground-based data, typically triplets or doublets are observed. In the latter case, the modulation peaks are detected on one side of the radial mode frequency (and of its harmonics). To claim the modulation we require at least two modulation side peaks corresponding to the same modulation frequency are detected in the frequency spectrum (see Sect.~\ref{ssec:modulation} for a more thorough discussion).

After prewhitening the first overtone and its harmonics, quite often the dominant signal in the frequency spectrum is unresolved with the just prewhitened 1O signal. It indicates that 1O is non-stationary; its amplitude and/or phase vary on a long timescale (of order of data length or higher). Such signals are often strong, dominate the frequency spectrum and increase the noise level, thereby hindering the detection of additional low-amplitude periodicities. To get rid of such non-stationary variation we employ time-dependent prewhitening as applied to {\it Kepler} space data by \cite{pam15} (see their Appendix A) and later applied to OGLE data eg. in \cite{ss16} or \cite{ns19}. In a nutshell, the data are divided into groups of specified length, $\delta t$, typically corresponding to single observing season or its half, and sine series is fitted independently to each group, keeping the frequencies fixed to that determined from all data fit. Only zero point, amplitudes, and phases of sine terms are adjusted. Such determined fits are then used to compute the residuals on a group-to-group basis. As a result amplitude or phase changes on the time scale longer than $\delta t$ are removed from the data, while possible faster changes, that can be well resolved, are still preserved. The technique can also be used to remove slow trends in the data when only group-dependent zero-points are subtracted from the data. After the time-dependent pre-whitening, the non-stationary signal(s) are removed, which reduces their amplitudes and the overall noise level and new low-amplitude signals are often detected.

\begin{figure}
\noindent\includegraphics[width=\columnwidth]{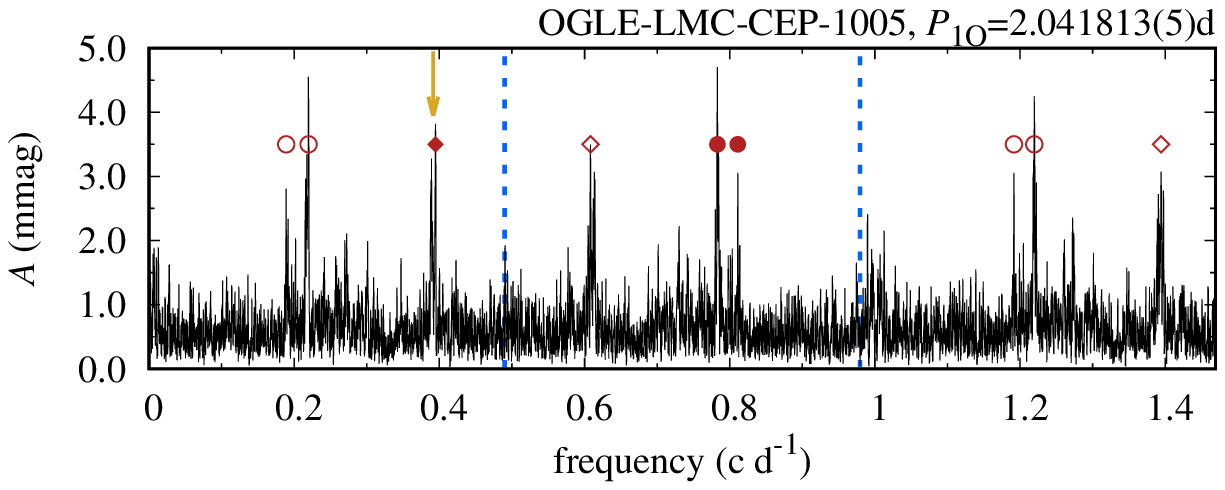}\\
\noindent\includegraphics[width=\columnwidth]{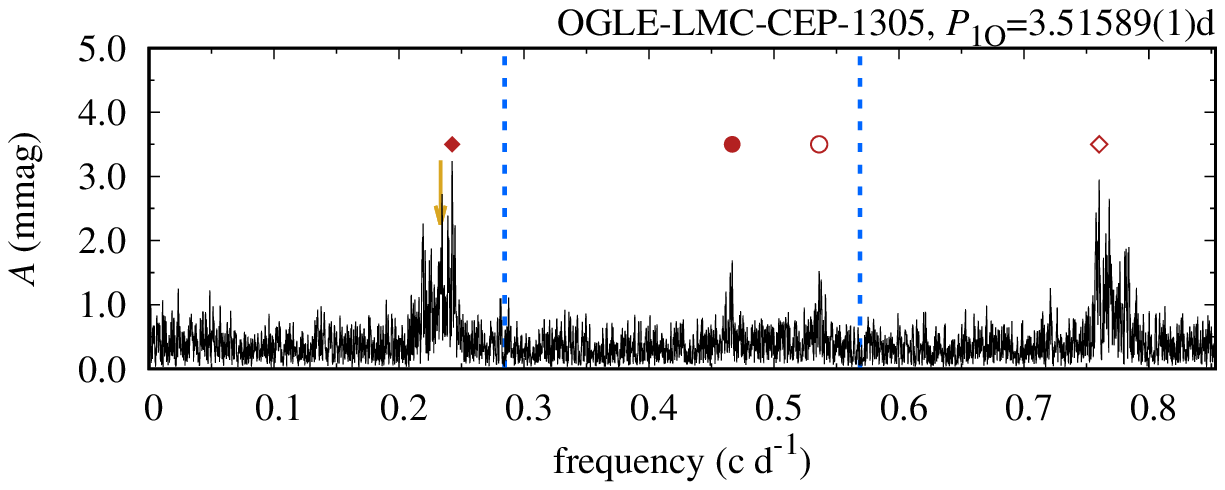}\\
\noindent\includegraphics[width=\columnwidth]{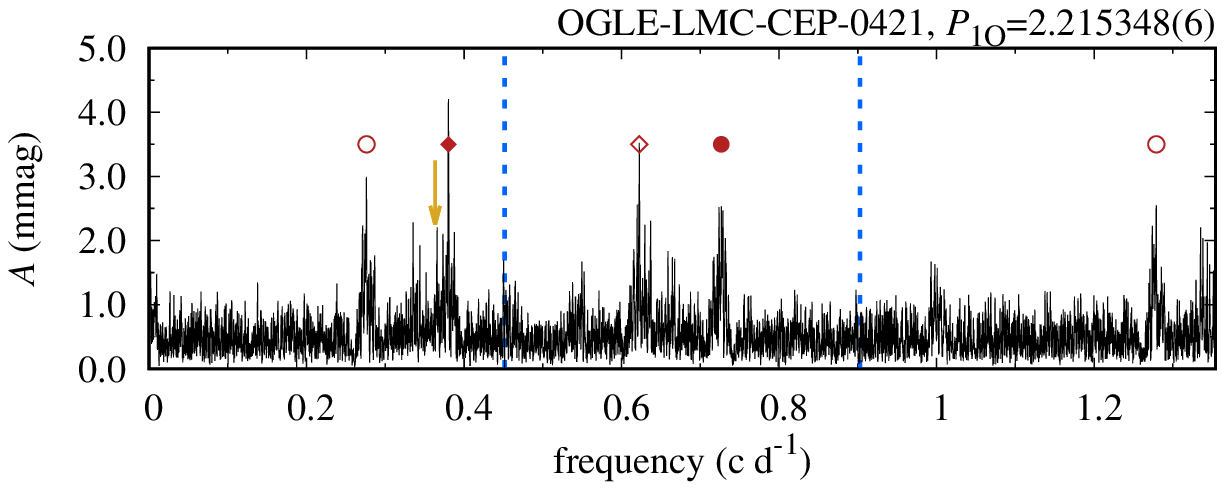}\\
\noindent\includegraphics[width=\columnwidth]{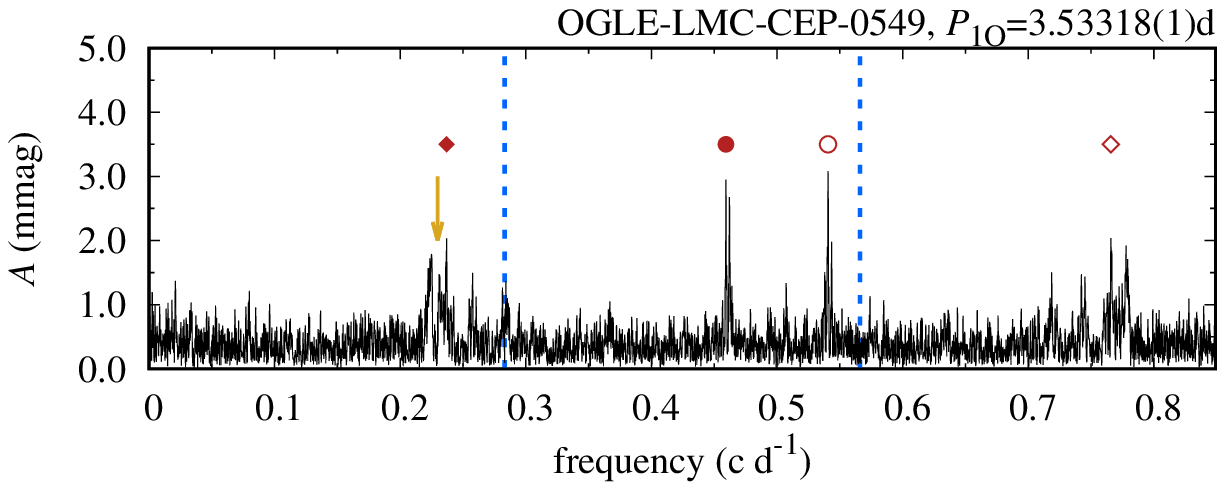}
\caption{Exemplary frequency spectra for analysed stars, after prewhitening with the first overtone frequency and its harmonics (dashed lines). Filled circles and diamonds mark the highest signals detected in the $\px/\pov=(0.60,\,0.65)$ range and corresponding sub-harmonic range, respectively. Open symbols mark locations of the corresponding daily aliases. Arrows indicate the exact location of the sub-harmonic of the highest signal detected in the $\px/\pov=(0.60,\,0.65)$ range.}
\label{fig:fspexample}
\end{figure}

Non-stationary variability may also manifest through more complex structures in the frequency spectrum, power excesses or wide bands of power. For Cepheids with period ratios $\px/\pov$ in the $(0.60,\,0.65)$ range, these take various forms, eg., bell-shaped power excess, U-shaped power excess, nearly rectangular bands of power excess, as already noted by \cite{ss16} for 1O Cepheids or by \cite{ns19} for RRc stars showing similar variability. In these stars a power excess centred at sub-harmonic frequency, $1/2\fx$, is quite often detected, as illustrated in Fig.~\ref{fig:fspexample} for a few stars ($1/2\fx$ is marked with an arrow). To characterise such signals, we follow the simplest approach adopted also in the just quoted studies: we include a single sine term in the solution with frequency corresponding to maximum amplitude within the power excess (filled diamond in Fig.~\ref{fig:fspexample}). Remnant and quite often resolved power is then present in the frequency spectrum after prewhitening. One would have to include several sine terms of close frequencies to remove such power excess. 

With OGLE data we face a common issue of ground-based data analysis: a possible confusion due to daily aliases. Aliases arise due to regularities in data sampling and their expected structure can be revealed by calculating the spectral window. OGLE data for Magellanic Clouds are seasonal, thus both daily and 1-year aliases are present and typically strong. For signals with $\sn\lesssim 6$ a few daily aliases may be of comparable height: sometimes a `comb' of four to six aliases with minute differences of the $\sn$ are present. Which alias corresponds to intrinsic stellar variability? The problem was faced eg., by \cite{om09} who studied RRc stars in $\omega$~Cen. For ambiguous cases, they provided two, or even three solutions with different interpretations (see their tab.~1). Two solutions were common: either a new double-periodic form of pulsation with period ratios $\px/\pov\approx 0.61$ or double-mode radial 1O+2O pulsation. With analysis of more data from various stellar systems, it became clear that the former explanation is correct \citep[see eg.][]{2009AcA....59....1S,2012MNRAS.424.2528S,2012A&A...540A..68C,2014A&A...570A.100S,pam15}. In this study, when selecting the intrinsic signal, whenever applicable, we make use of the prior knowledge on various forms of double-periodic pulsation among 1O Cepheids. This is illustrated in Fig.~\ref{fig:fspexample}, with the frequency spectra for \idlmc{0421} and \idlmc{0549}\footnote{In the following full OGLE IDs, eg., OGLE-LMC-CEP-0421, are abbreviated as \idlmc{0421}}.

For \idlmc{0421}, after prewhitening with first overtone frequency and its harmonic, the highest $\sn=7.7$ signal is detected at $\nu=0.380\cd$ (marked with filled diamond). After prewhitening with this signal two daily aliases come to attention. The higher $\sn=5.6$ signal is located at low-frequencies (open circle at $\nu=0.276\cd$) and the lower $\sn=4.9$ signal at higher frequency (filled circle at $\nu=0.727\cd$). Since the latter signal corresponds to $P/\pov=0.621$ and we already detected a power excess centred at its sub-harmonic (with the highest signal at $\nu=0.380\cd$), we consider this alias a true periodicity. We include it in our solution and consider the star as a classical example of double-periodic star with period ratio $\px/\pov$ in the $(0.60,\,0.65)$ range. It nicely fits the middle sequence in the Petersen diagram, and has a signal centred at sub-harmonic frequency present.

For \idlmc{0549}, after prewhitening with first overtone frequency and its harmonics, three daily aliases come to attention. The highest $\sn=7.7$ signal is located at $\nu=1.462\cd$ (not visible in Fig.~\ref{fig:fspexample}), then there is a $\sn=7.4$ signal at $\nu=0.541\cd$ (open circle) and a $\sn=7.0$ signal at $\nu=0.459\cd$ (filled circle). Since for the last signal we find $P/\pov=0.616$ and we can clearly see a signal centred at its sub-harmonic (arrow) we consider it a true periodicity, even thou it is the third highest signal in a 1-day alias family.

The above outlined procedure is subjective and in some cases may lead to identification of false periodicities that are, in fact, aliases of the correct ones. This is mitigated by conservative choices and selecting the lower alias only when $\sn$ differences are small and there are good arguments behind (like given in the above examples). Error is still possible, but rather for a few Cepheids out of hundreds with additional signals considered. Consequently, conclusions about statistical properties of the studied groups of double-periodic pulsation are robust.

\section{Results}
\label{sec:results}

\subsection{Overview}

OGLE-IV photometric data on 1766 1O Cepheids from the LMC and 1784 1O Cepheids from the SMC were analysed. Of these, in 45\,per cent of stars in the SMC and in 25\,per cent of stars in the LMC, no significant additional signal beyond radial mode and its harmonics was detected in automatic procedure outlined in the previous section.

\begin{figure*}
\includegraphics[width=\columnwidth]{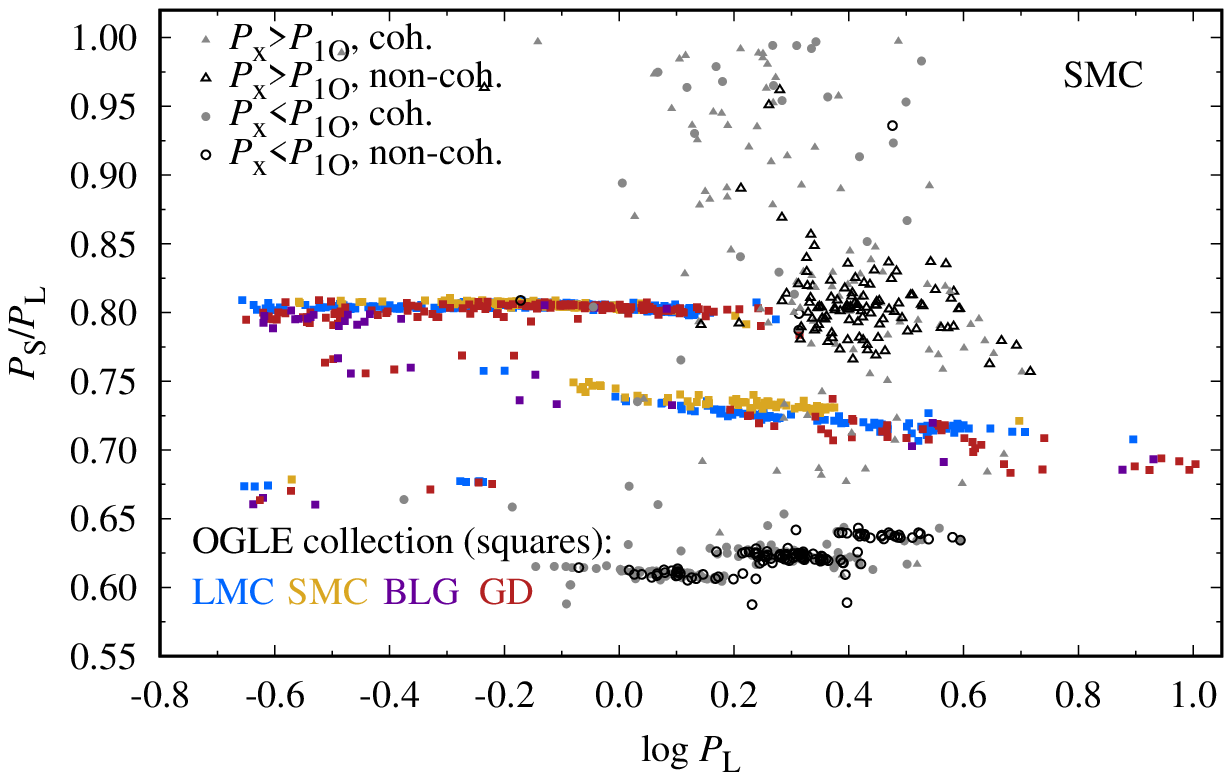}
\includegraphics[width=\columnwidth]{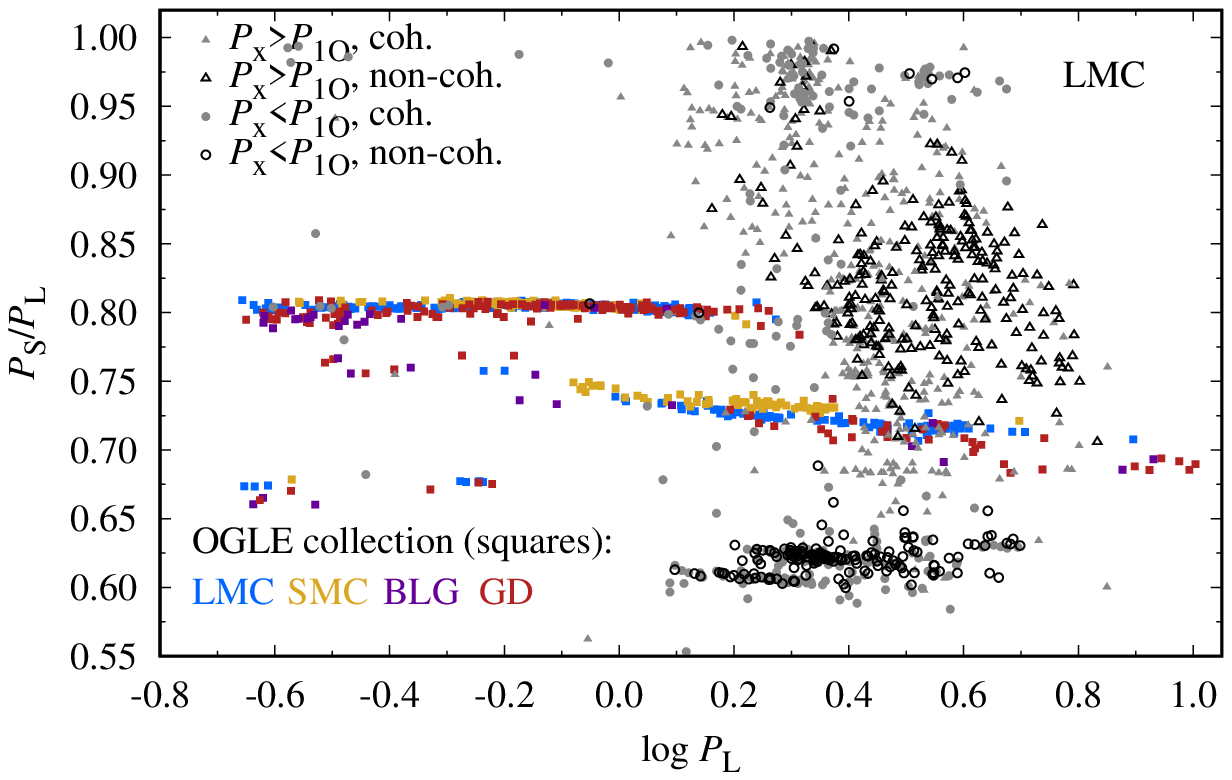}
\caption{The Petersen diagrams (ratio of shorter to longer period, $P_{\rm S}/P_{\rm L}$, vs. the longer period, $P_{\rm L}$) for the SMC (left) and the LMC (right) with period ratios for all additional periodicities detected in the analysis, except those that can be interpreted as due to periodic modulation. If $\px>\pov$, then $\pov/\px$ vs. $\px$ is plotted with triangles, while for $\px<\pov$, $\px/\pov$ vs. $\pov$ is plotted with circles. Filled, dark gray symbols correspond to coherent additional variability, while open black symbols are used when unresolved power or power excess is detected in the frequency spectrum after prewhitening the data with additional periodicity. Data on radial double-mode and triple-mode Cepheids from OGLE-CVS are plotted with small colored squares (different colors for different stellar systems: LMC, SMC, Galactic bulge, BLG, and Galactic disk, GD, to provide context.}
\label{fig:petall}
\end{figure*}

Additional signals were detected in over thousand 1O Cepheids in both Clouds. Corresponding period ratios (shorter to longer) are plotted in the Petersen diagrams in Fig.~\ref{fig:petall}, separately for SMC (left) and LMC (right). Triangles and circles correspond, respectively, to additional periodicities with period longer than 1O period and shorter than 1O period. Thus, either $\pov/\px$ vs. $\px$ (triangles), or $\px/\pov$ vs. $\pov$ (circles) is plotted.

While the number of 1O Cepheids in both Magellanic Clouds is similar, additional periodicities are much more frequent in the LMC. We discuss it in detail in Sect.~\ref{ssec:unclassified}. Here we just note that since SMC is further away than LMC, its Cepheids are dimmer, and the detection limit is higher. It seems, however, that higher incidence rate of stars with additional periodicities is an intrinsic property of the LMC sample, at least for some range of first overtone periods. 

Based on the location in the Petersen diagram, we first classify the stars into a few already known classes of double-periodic Cepheids. These include three classes of double-mode radial pulsation, discussed in Sect.~\ref{ssec:dm}, double-periodic pulsation with period ratios, $\px/\pov$, in the $(0.60,\,0.65)$ range (Sect.~\ref{ssec:61}) and double-periodic pulsation with period ratios $\pov/\px$ clustered around $0.684$ (Sect.~\ref{ssec:686}). Then we discuss the remaining double-periodic variables in Sect.~\ref{ssec:unclassified}. Stars in which modulation side peaks were detected are discussed in Sect.~\ref{ssec:modulation}. In several stars multiple forms of double-periodic pulsation are present simultaneously.


\subsection{Double-Mode Radial Pulsation}\label{ssec:dm}

To identify the candidates for double-mode radial pulsation, we compare the location of double-periodic stars in the Petersen diagram, with the location of known double-mode radial pulsators from the OGLE-CVS in the LMC, SMC, Galactic bulge and Galactic disc \citep[][]{o4_clouds,o4_gal2}. In Fig.~\ref{fig:petall}, these stars are plotted with small filled squares, colour-coded according to the stellar system. 1O+2O stars from the OGLE-CVS form a rather tight progression with $\poov/\pov$ centred at $\approx 0.8$. The more significant scatter may be noted at the short and long-period ends of this progression. Metallicity dependence is weak, as may be inferred from comparison of data for the four stellar systems. Progression for F+1O stars is slanted, with $\pov/\pf$ decreasing from $\approx0.76$ at the short-period end, to $\approx0.69$ at the long-period end. As in the case of 1O+2O stars, higher dispersion of period ratios is visible at short and long-period ends of the progression. Metallicity dependence is clear as best visible comparing the progressions for the most numerous LMC and SMC samples. 1O+3O radial double-mode pulsation is scarce; interestingly, this combination of radial modes is more frequent in the triple-mode pulsators than in the double-mode pulsators. Data for both groups are used in Fig.~\ref{fig:petall}. The 1O+3O stars are clustered in two groups with a rather large spread of $\pooov/\pov$, in between $0.66$ and $0.68$. 

\begin{figure}
\includegraphics[width=\columnwidth]{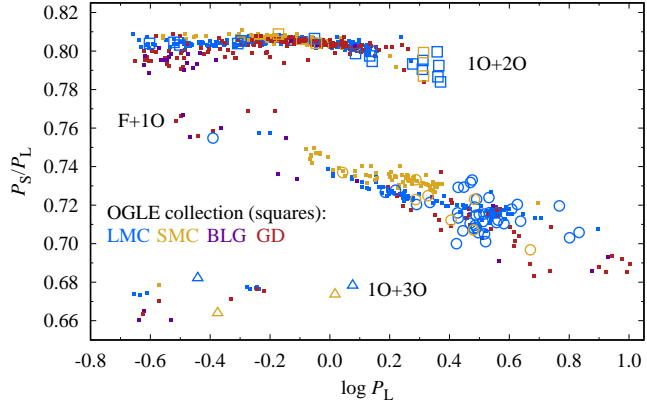}
\caption{The Petersen diagram with new candidates for double-mode radial pulsation marked with large open symbols. Small filled squares represent data on the double-mode and triple-mode pulsators from the OGLE-CVS in different stellar systems (indicated with different colors as described inside the plot).}
\label{fig:petrad}
\end{figure}

In Fig.~\ref{fig:petrad}, we show new candidates for double-mode radial pulsators, discussed in more detail in the following sections. 

\subsubsection{Candidates for F+1O double-mode radial pulsation}\label{sssec:f1o}

In Fig.~\ref{fig:petall}, several new candidates for double-mode radial F+1O pulsation may be identified. These are stars for which $\px>\pov$, that fall along the progression of F+1O double-mode radial pulsators from the OGLE-CVS. 38 candidates were identified in the LMC and 7 candidates were identified in the SMC. Data for these new candidates are given in Tab.~\ref{tab:f1o}. In the consecutive columns we report: star's ID, first overtone period, fundamental mode period, period ratio, $\pov/\pf$, amplitudes of the first overtone, $\aov$, and of the fundamental, $\af$, modes, signal-to-noise for the detection of fundamental mode and remarks including information on the detected combination frequencies. In Fig.~\ref{fig:petrad}, we show the location of these candidates in the Petersen diagram for double-mode radial pulsation (open circles). We note that for most of our candidates, period of the fundamental mode is longer than is typical for double-mode F+1O pulsators in the OGLE-CVS. While majority of candidates in the LMC/SMC fall within the respective progression formed by the F+1O OGLE-CVS stars, or along its extension, for several stars period ratios are larger, or lower than typical in a given system. This implies that metallicity of these stars may be lower/higher than typical for Cepheids in a given system.

Amplitude ratios are usually very low; the median ratio, $\af/\aov$, is 2.5\, per cent. The most extreme case in which amplitude ratio is 44\, per cent (\idlmc{3362}) is discussed below.

%

\begin{table*}
\caption{Candidates for new double-mode F+1O radial pulsators. Consecutive columns contain star's ID, first overtone period, fundamental mode period, period ratio, $\pov/\pf$, amplitudes of the first overtone, $\aov$, and of the fundamental, $\af$, modes, signal-to-noise ratio for the detection of fundamental mode. Remarks in the last column can be the following: `nsO' - non stationary dominant (1O) variability, `nsX' -- non stationary additional variability, `tdp' -- additional signal firmly detected only after the time-dependent prewhitening, `al' a lower alias was selected, `0.61' -- additional variability with $\px/\pov\in(0.60,\,0.65)$ was detected, `0.68' -- additional variability with $\pov/\px\approx 0.684$ was detected, `ap' -- additional variability that does not fit the `0.61' and `0.68' schemes was detected in the star.}
\label{tab:f1o}
\begin{tabular}{lrrrrrrr}
\hline
Star & $\pov$\,(d) & $\pf$\,(d) &  $\pov/\pf$ & $\aov$ (mag) & $\af$ (mag) & $\sn$    & Additional frequencies \& Remarks\\
\hline
\idlmc{0010} & 2.565580(6) & 3.6086(7)  & 0.7110 & 0.1268 & 0.0024 & 4.1 & ap\\  
\idlmc{0243} & 2.343337(6) & 3.2759(5)  & 0.7153 & 0.1189 & 0.0029 & 5.2 & 0.61, nsO, nsX\\
\idlmc{0297} & 4.80681(2)  & 6.813(2)   & 0.7059 & 0.1071 & 0.0016 & 4.4 & 0.61, nsO, nsX, tdp, al \\
\idlmc{0342} & 2.193763(5) & 3.0511(5)  & 0.7190 & 0.0882 & 0.0017 & 4.2 & \\
\idlmc{0583} & 2.147148(4) & 3.0332(3)  & 0.7079 & 0.0920 & 0.0023 & 6.5 & $\ff+\fmodul$, $\ff+2\fmodul$ \\
\idlmc{0839} & 2.193277(5) & 2.9919(4)  & 0.7331 & 0.1212 & 0.0030 & 5.2 & nsO, nsX, 0.61 \\  
\idlmc{0868} & 2.160823(3) & 2.9528(5)  & 0.7318 & 0.0960 & 0.0025 & 4.4 & nsO, 0.61, ap \\ 
\idlmc{1134} & 3.09037(1)  & 4.3426(9)  & 0.7116 & 0.0819 & 0.0025 & 4.4 & nsO \\
\idlmc{1217} & 3.05071(3)  & 4.2350(9)  & 0.7204 & 0.0318 & 0.0017 & 4.5 & nsO, nsX \\
\idlmc{1224} & 1.912693(7) & 2.6714(4)  & 0.7160 & 0.0722 & 0.0025 & 4.1 & nsO, ap, al\\ 
\idlmc{1312} & 2.258804(4) & 3.1587(5)  & 0.7151 & 0.0783 & 0.0014 & 4.4 & \\
\idlmc{1346} & 2.462782(9) & 3.4021(6)  & 0.7239 & 0.1101 & 0.0018 & 4.3 & 0.61, 0.684, ap, nsO, tdp\\
\idlmc{1425} & 2.713019(6) & 3.8192(4)  & 0.7104 & 0.1209 & 0.0025 & 7.4 & $\fov+\ff$, nsO, 0.61, ap \\ 
\idlmc{1643} & 2.597817(5) & 3.6474(5)  & 0.7122 & 0.0969 & 0.0019 & 5.6 & 0.61, ap \\
\idlmc{1747} & 2.186281(3) & 3.0866(3)  & 0.7083 & 0.1084 & 0.0028 & 7.6 & nsO, ap\\ 
\idlmc{1918} & 1.404156(2) & 1.9494(2)  & 0.7203 & 0.0832 & 0.0018 & 4.2 & \\
\idlmc{2137} & 4.21030(2)  & 5.8508(9)  & 0.7196 & 0.0785 & 0.0025 & 8.1 & 0.684 \\
\idlmc{2179} & 1.967818(4) & 2.7847(4)  & 0.7067 & 0.1098 & 0.0024 & 4.9 & nsO, 0.61, ap\\
\idlmc{2357} & 2.324088(7) & 3.3153(5)  & 0.7010 & 0.1097 & 0.0019 & 4.6 & nsO, tdp, 0.61, ap\\ 
\idlmc{2406} & 2.091337(4) & 2.9431(4)  & 0.7106 & 0.0926 & 0.0019 & 4.5 & 0.684, ap\\  
\idlmc{2409} & 2.318955(4) & 3.2890(4)  & 0.7051 & 0.1159 & 0.0021 & 5.2 & nsO, ap\\
\idlmc{2540} & 2.911084(8) & 4.0548(9)  & 0.7179 & 0.0951 & 0.0013 & 4.2 & nsO, tdp, 0.61\\
\idlmc{2554} & 2.581272(4) & 3.6051(4)  & 0.7160 & 0.1296 & 0.0028 & 6.3 & 0.61, ap \\
\idlmc{2590} & 1.205806(2) & 1.6569(1)  & 0.7278 & 0.0829 & 0.0026 & 4.6 & nsO, tdp\\
\idlmc{2617} & 2.045449(3) & 2.8046(3)  & 0.7293 & 0.1095 & 0.0020 & 5.4 & 0.61 \\
\idlmc{2685} & 2.170927(4) & 3.0596(3)  & 0.7096 & 0.0965 & 0.0023 & 5.3 & nsO, nsX, 0.61, ap\\
\idlmc{2719} & 2.447373(6) & 3.4178(6)  & 0.7161 & 0.0855 & 0.0017 & 4.2 & 0.61, ap\\ 
\idlmc{2732} & 1.958648(4) & 2.6859(4)  & 0.7292 & 0.0996 & 0.0016 & 4.0 & 0.61, ap\\ 
\idlmc{2822} & 2.214104(6) & 3.1376(3)  & 0.7057 & 0.1004 & 0.0044 & 7.1 & nsO, 0.61, ap \\ 
\idlmc{2845} & 2.312868(4) & 3.2508(5)  & 0.7115 & 0.1371 & 0.0023 & 4.0 & ap\\  
\idlmc{3271} & 1.855232(3) & 2.6503(3)  & 0.7000 & 0.1037 & 0.0040 & 5.8 & \\
\idlmc{3321} & 4.45108(2)  & 6.331(2)   & 0.7030 & 0.1076 & 0.0030 & 4.9 & 0.61, ap \\
\idlmc{3362} & 0.3069825(4)& 0.406681(2)& 0.7548 & 0.0804 & 0.0354 &12.8 & $\fov+\ff$, $\fov-\ff$  \\
\idlmc{3606} & 1.114184(2) & 1.53331(8) & 0.7267 & 0.0955 & 0.0041 & 5.3 & nsO \\
\idlmc{3669} & 2.237794(8) & 3.0948(4)  & 0.7231 & 0.1180 & 0.0045 & 4.8 & \\
\idlmc{4037} & 1.916503(4) & 2.6867(3)  & 0.7133 & 0.1005 & 0.0024 & 4.1 & 0.61, ap \\
\idlmc{4370} & 2.200759(7) & 3.0793(5)  & 0.7147 & 0.0944 & 0.0027 & 4.5 & nsO, ap \\
\idlmc{4624} & 1.559273(5) & 2.1614(3)  & 0.7214 & 0.0745 & 0.0027 & 3.8 & 0.684, al \\
\hline
\idsmc{0213} & 1.809871(4) & 2.5412(2)  & 0.7122 &  0.1123 & 0.0047 & 6.3 & \\ 
\idsmc{1472} & 1.421955(5) & 1.9395(2)  & 0.7332 &  0.0978 & 0.0039 & 4.5 & nsO, tdp \\ 
\idsmc{1475} & 3.26095(5)  & 4.680(1)   & 0.6968 &  0.090  & 0.0023 & 4.2 & nsO, tdp \\ 
\idsmc{1961} & 0.812985(1) & 1.10319(6) & 0.7369 &  0.100  & 0.005  & 4.2 & \\  
\idsmc{3453} & 2.201674(4) & 3.0451(4)  & 0.7230 &  0.1121 & 0.0021 & 4.2 & ap \\ 
\idsmc{3977} & 1.542793(4) & 2.1282(2)  & 0.7249 &  0.1206 & 0.0036 & 4.6 & nsO, tdp, 0.61 \\
\idsmc{4845} & 1.399198(4) & 1.9355(2)  & 0.7229 &  0.1014 & 0.0040 & 4.1 & al  \\ 
\idsmc{4890} & 2.136932(5) & 3.0223(3)  & 0.7071 &  0.1276 & 0.0046 & 5.5 & \\
\hline
\end{tabular}
\end{table*}


{\it Remarks on individual stars:}

\idlmc{0583}. The candidate for radial F mode, detected with $\sn=6.5$, appears to be modulated with $\pmodul=90.3(2)$\,d, as two multiplet components are detected at $\ff+\fmodul$ and $\ff+2\fmodul$ (of higher amplitude; $\fmodul=1/\pmodul$). After prewhitening, a weak ($\sn=3.2$), but distinct signal is present at $\ff+3\fmodul$. Since all four equidistant signals (including $\ff$) are of $\sim$comparable amplitude, we cannot firmly classify this star, and conclude about possible modulation of specific frequency.

\idlmc{2137}. F-mode is prominent ($\sn=8.1$), but even more prominent is additional signal detected at $\pov/\px=0.6856$. There are three more candidates (with weaker detection of the radial F-mode) that show additional signal of the same category. These stars may be important to understand the nature of signals centred at $\pov/\px\approx 0.684$, as discussed in more detail in Sect.~\ref{ssec:686}.

\idlmc{3362}. F-mode is prominent ($\sn=12.8$); its amplitude constitutes 44\,per cent of the first overtone amplitude, which is the highest amplitude ratio in our sample. On the other hand, first overtone amplitude is among the lowest in the considered sample of F+1O candidates and both pulsation periods are the shortest in the analysed sample and among the shortest for F+1O Cepheids in the OGLE-CVS. In fact, the short-period end of the F+1O Cepheid sequence overlaps with the long-period end of the F+1O high amplitude delta Scuti stars (HADS); the discussed star fits well both sequences and so is in the transition region between HADS and classical Cepheids.


\subsubsection{Candidates for 1O+2O double-mode radial pulsation}\label{sssec:1o2o}

Inspection of Petersen diagram in Fig.~\ref{fig:petall} reveals several stars with $\px<\pov$, that fall along the progression of 1O+2O double-mode radial pulsators from OGLE-CVS, or along its extension. Based on the location in the Petersen diagram, we classify 17 stars from the LMC and 4 stars from the SMC as new candidates for double-mode radial 1O+2O pulsation. Basic data for these stars are collected in Tab.~\ref{tab:1o2o} and they are highlighted in the Petersen diagram in Fig.~\ref{fig:petrad} with large open squares. In all but a few stars amplitude of the supposed 2O is small and the detections are weak, typically with $4<\sn<5$, and only in a few cases $\sn$ exceeds 5. The median amplitude ratio, $\aoov/\aov$, is 5\,per cent; the highest amounts to 34\,per cent and corresponds to a star discussed below.

{\it Remarks on individual stars:}

\begin{figure}
\includegraphics[width=\columnwidth]{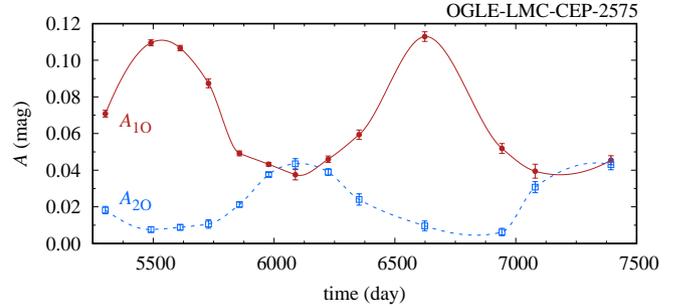}
\caption{Variation of amplitudes of the 1O and 2O modes in \idlmc{2575} extracted with the time-dependent Fourier analysis. Time is $\mathrm{HJD}-245\,0000$.}
\label{fig:2575}
\end{figure}

The most prominent and interesting case is \idlmc{2575} in which second overtone is firmly detected with $\sn=12.5$. Both frequencies are non-stationary; unresolved or barely resolved signals are detected both at $\fov$ and $\foov$ after prewhitening. In fact, clear, large amplitude modulation is well visible just by inspecting the photometric data, and a star is listed as having the Blazhko modulation in the remarks file of the OGLE-CVS \citep{o4_clouds_multimode}. A procedure of time-dependent prewhitening allows us to detect further significant frequencies, in particular, the linear combination $\fov+\foov$, and to study the variation of 1O and 2O amplitudes, visualised in Fig.~\ref{fig:2575}. The amplitude changes are clearly anti-correlated. We conclude that \idlmc{2575} is another member of the group of double-overtone Cepheids, modulated on long time scales with anti-correlated amplitude changes, studied in detail by \cite{mk09}.

Another interesting star is \idlmc{0071} as additional signal with $\px/\pov\in(0.60,\,0.65)$ is detected (see Sect.~\ref{ssec:61}) in addition to the dominant 1O and candidate 2O. Thus, this star may be a promising target for asteroseismic modelling. Admittedly the detections of additional signals in this star are weak ($\sn<5$).

\begin{table*}
\caption{Candidates for new double-mode 1O+2O radial pulsators. Structure of the table and remarks in the last column are the same as in Tab.~\ref{tab:f1o}.}
\label{tab:1o2o}
\begin{tabular}{lrrrrrrr}
\hline
Star & $\pov$\,(d) & $\poov$\,(d) &  $\poov/\pov$ & $\aov$ (mag) & $\aoov$ (mag) & $\sn$ & Additional frequencies \& Remarks\\
\hline
\idlmc{0071} & 1.385893(2) & 1.10130(6) & 0.7947 & 0.1489 & 0.0029 &  4.2 & 0.61\\
\idlmc{0232} & 0.301231(1) & 0.242359(4)& 0.8046 & 0.0358 & 0.0093 &  4.3 & \\
\idlmc{0302} & 2.343800(7) & 1.8375(2)  & 0.7840 & 0.0710 & 0.0019 &  4.1 & \\
\idlmc{1334} & 2.31368(1)  & 1.83350(9) & 0.7925 & 0.0440 & 0.0030 &  7.2 & ap\\
\idlmc{1345} & 2.298207(7) & 1.8078(2)  & 0.7866 & 0.0608 & 0.0016 &  4.5 & \\
\idlmc{1435} & 0.2495233(3)& 0.200612(2)& 0.8040 & 0.0909 & 0.0095 &  4.3 & \\
\idlmc{1662} & 0.3151630(4)& 0.253116(3)& 0.8031 & 0.0640 & 0.0059 &  4.6 & \\
\idlmc{1710} & 0.888527(3) & 0.71657(2) & 0.8065 & 0.0759 & 0.0076 &  5.8 & nsO, nsX\\
\idlmc{2575} & 1.37710(1)  & 1.10146(2) & 0.7998 & 0.0677 & 0.0231 & 12.5 & $\fov+\foov$, nsO, nsX, Blazhko\\
\idlmc{2872} & 2.038338(5) & 1.6113(1)  & 0.7905 & 0.1013 & 0.0029 &  4.4 & nsO\\
\idlmc{3230} & 1.892812(6) & 1.5016(1)  & 0.7933 & 0.0603 & 0.0019 &  4.0 & ap \\
\idlmc{3390} & 0.492180(1) & 0.395665(6)& 0.8039 & 0.0813 & 0.0120 &  4.6 & \\ 
\idlmc{3519} & 0.5050271(5)& 0.406690(6)& 0.8053 & 0.1290 & 0.0080 &  5.1 & nsO\\
\idlmc{3672} & 1.359207(3) & 1.08395(4) & 0.7975 & 0.1162 & 0.0061 &  6.4 & nsO, ap\\ 
\idlmc{3945} & 1.218720(2) & 0.97340(4) & 0.7987 & 0.0944 & 0.0032 &  5.0 & nsO, tdp\\
\idlmc{4391} & 2.285239(7) & 1.8273(1)  & 0.7996 & 0.0777 & 0.0017 &  5.0 & nsO, tdp\\
\idlmc{4565} & 2.046588(3) & 1.6272(1)  & 0.7951 & 0.1344 & 0.0024 &  4.6 & al\\
\hline
\idsmc{1003} & 0.901880(1) & 0.72517(2) & 0.8041 & 0.164  & 0.008  &  5.5 & $\fov+\foov$\\ 
\idsmc{2209} & 2.055730(4) & 1.6180(1)  & 0.7870 & 0.1397 & 0.0029 &  4.6 & al, nsX\\ 
\idsmc{2988} & 2.056799(7) & 1.6437(1)  & 0.7992 & 0.0777 & 0.0034 &  5.0 & nsX\\ 
\idsmc{4881} & 0.673569(2) & 0.54479(1) & 0.8088 & 0.130  & 0.018  &  6.7 & nsX, al\\ 
\hline
\end{tabular}
\end{table*}


\subsubsection{Candidates for 1O+3O double-mode radial pulsation}\label{sssec:1o3o}

Four candidates for 1O+3O radial double-mode pulsation are reported in Tab.~\ref{tab:1o3o}: two stars in the LMC and two in the SMC. The additional signal in \idlmc{2561} is detected only after time-dependent prewhitening is applied due to non-stationary nature of 1O. In all stars the detections are weak ($\sn\lesssim5.0$), the additional signals are coherent and we do not detect any combination frequencies with 1O. Interestingly, in \idsmc{4092}, additional signals are detected with $\pov/\px\in (0.6,\, 0.65)$ (see Sect.~\ref{ssec:61}).

In Fig.~\ref{fig:petrad}, location of the 1O+3O candidates is plotted in the Petersen diagram (open triangles) and may be compared with $\pooov/\pov$ ratios reported in the double-mode (1O+3O) and in the triple-mode (1O+2O+3O) pulsators reported by OGLE in the LMC, SMC, Galactic bulge and Galactic disk. The 1O+3O stars already reported in the OGLE-CVS are clustered in two groups, around $\log\pov=-0.6$ and $\log\pov=-0.25$. Two new candidates are located in between these groups, while two other have a much longer pulsation period ($\log\pov\approx0.08$).
 
\begin{table*}
\caption{Candidates for new double-mode 1O+3O radial pulsators. Structure of the table and remarks in the last column are the same as in Tab.~\ref{tab:f1o}.}
\label{tab:1o3o}
\begin{tabular}{lrrrrrrr}
\hline
Star & $\pov$\,(d) & $\pooov$\,(d) &  $\pooov/\pov$ & $\aov$ (mag) & $\aooov$ (mag) & $\sn$    & Remarks\\
\hline
\idlmc{1274} & 0.3616755(8)& 0.246708(3) & 0.6821 & 0.052  & 0.007  & 4.2 & \\
\idlmc{2561} & 1.192038(2) & 0.80854(3)  & 0.6783 & 0.0802 & 0.0026 & 4.6 & nsO, tdp\\
\hline
\idsmc{4092} & 1.0409489(7)& 0.70124(2)  & 0.6737 & 0.1811 &  0.0037 & 5.1 & 0.61\\
\idsmc{4465} & 0.4216633(6)& 0.279966(4) & 0.6640 & 0.132  &  0.010  & 4.4 & \\

\hline
\end{tabular}
\end{table*}

\subsection{Double-periodic stars with $\px/\pov$ in between 0.60 and 0.65}\label{ssec:61}

Classical Cepheids with the dominant radial 1O and additional signals with $\px/\pov\in(0.60,\,0.65)$, were first reported by \cite{mk08} and \cite{o3_lmc_cep}, based on the analysis of OGLE photometry in the LMC. Later 138 1O Cepheids of the same type were detected in the SMC \citep{o3_smc_cep}. In the LMC, two clear and close sequences were reported in the Petersen diagram, while for the SMC three sequences could be distinguished. The first in-depth analysis of this form of pulsation for 138 SMC stars was presented in \cite{ss16}. It was found that signals detected at $\fx$ are often accompanied with signals centred, or located close to the sub-harmonic, $1/2\fx$. The latter signals are often non-coherent: wide bands of power excesses are detected. While the power excess is usually well centred at $1/2\fx$, the highest signal within, that had been used to characterise this structure (its frequency denoted $\fsh$), may be a bit off. 

In Tab.~A1 in the on-line Appendix, section of which is shown in Tab.~\ref{tab:61} for a reference, we collect the properties of additional signals with $\px/\pov\in(0.60,\,0.65)$ for the SMC (225 stars; $12.6$\,per cent of the sample) and the LMC (291 stars, $16.5$\,per cent of the sample). The consecutive columns contain star's id, first overtone frequency, $\fov$, frequency of the additional signal, $\fx$, corresponding period ratio, $\px/\pov$, amplitude of the 1O, amplitude ratio, $\ax/\aov$, $\sn$ for the detection at $\fx$ and remarks (described in the Table's caption). 

For stars in which signals centred at $1/2\fx$ were also detected, additional data are included in Tab.~A2 in the on-line Appendix, section of which is shown in Tab.~\ref{tab:61sh} for a reference. The consecutive columns contain star's id, period ratio, $\px/\pov$, frequency of the additional signal, $\fx$, frequency of the highest signal close to sub-harmonic of $\fx$, $\fsh$, amplitude of the additional signal, $\ax$, amplitude ratio, $\ash/\ax$, $\sn$ for the detection at $\fsh$ and remarks (described in the Table's caption).

\begin{table*}
\caption{Properties of first overtone Cepheids with period ratios, $\px/\pov$, in the $(0.60,\,0.65)$ range. Consecutive columns contain: star's id, first overtone frequency, $\fov$, frequency of the additional variability, $\fx$ , corresponding period ratio, $\px/\pov=\fov/\fx$, amplitude of the first overtone, $\aov$, and amplitude ratio, $\ax/\aov$, and remarks: `al' -- daily alias of signal at $\fx$ is higher; `nsX' -- complex appearance of the signal at $\fx$; `nsO' -- non-stationary first overtone; `cf' -- combination frequency of $\fx$ and $\fov$ detected; `sh' -- power excess at sub-harmonic frequency (around $1/2\fx$) detected; `ap'-- additional periodicity detected; `tdp' -- time-dependent analysis was conducted.}
\label{tab:61}
\begin{tabular}{lrrrrrrl}
\hline
Star & $\fov$ (d$^{-1}$) & $\fx$ (d$^{-1}$) & $\px/\pov$ & $\aov$ (mag) & $\ax/\aov$ & $\sn$ & Remarks\\
\hline
OGLE-SMC-CEP-0088 & 0.650005(1)  & 1.03295(5)  & 0.6293 & 0.1521(7) & 0.024 &  4.4 & al \\ 
OGLE-SMC-CEP-0212 & 0.574370(1)  & 0.91905(6)  & 0.6250 & 0.1026(6) & 0.043 &  5.5 & sh, nsX, cf \\ 
OGLE-SMC-CEP-0251 & 0.5565345(9) & 0.89055(4)  & 0.6249 & 0.1376(5) & 0.022 &  4.5 & sh, al\\ 
OGLE-SMC-CEP-0260 & 0.887240(1)  & 1.45102(3)  & 0.6115 & 0.1564(7) & 0.034 &  5.8 & cf, al \\ 
OGLE-SMC-CEP-0280 & 0.596946(1)  & 0.95146(4)  & 0.6274 & 0.1378(6) & 0.026 &  4.3 & sh, ap \\ 
\ldots            &              &             &        &           &       &      &\\
\hline
\end{tabular}
\end{table*}

\begin{table*}
\caption{Stars with significant power excess centred at sub-harmonic frequency, $1/2\fx$. Consecutive columns contain: star's id, period ratio, $\px/\pov$,
frequency of the additional variability, $\fx$ , frequency of the highest peak detected around $1/2\fx$, $\fsh$, frequency ratio, $\fsh/\fx$, amplitude of the additional
variability, $\ax$, and amplitude ratio, $\ash/\ax$, approximate $\sn$ for the peak at $\fsh$ and remarks: `nss' -- complex appearance of the signal at $\fsh$; `nss-broad' -- particularly broad power excess at $\fsh$;
 `al' -- daily alias of signal at $\fsh$ is higher;
`tdp' -- time-dependent pre-whitening of all signals except $\fsh$ conducted.}
\label{tab:61sh}
\begin{tabular}{lrrrrrrrl}
\hline
Star & $\px/\pov$ & $\fx$ (d$^{-1}$) & $\fsh$ (d$^{-1}$) & $\fsh/\fx$ & $\ax$ (mag) & $\ash/\ax$ & $\sn$ & Remarks\\
\hline
OGLE-SMC-CEP-0212 & 0.6250 & 0.91905(6)  & 0.45285(4) & 0.4927  & 0.0044(6) & 0.82 &  4.7 & nss, al \\ 
OGLE-SMC-CEP-0251 & 0.6249 & 0.89055(4)  & 0.44279(4) & 0.4972  & 0.0031(5) & 1.30 &  5.5 & nss, al \\ 
OGLE-SMC-CEP-0280 & 0.6274 & 0.95146(4)  & 0.49075(4) & 0.5158  & 0.0035(6) & 1.14 &  4.8 & \\ 
OGLE-SMC-CEP-0348 & 0.6208 & 0.77940(4)  & 0.38365(3) & 0.4922  & 0.0034(6) & 1.33 &  6.3 & nss-broad, al \\ 
OGLE-SMC-CEP-0477 & 0.6344 & 0.40006(4)  & 0.19209(4) & 0.4802  & 0.0027(5) & 1.45 &  5.0 & nss, al\\ 
\ldots            &        &             &            &         &           &       &      &\\
\hline
\end{tabular}
\end{table*}

\begin{figure*}\includegraphics[width=\columnwidth]{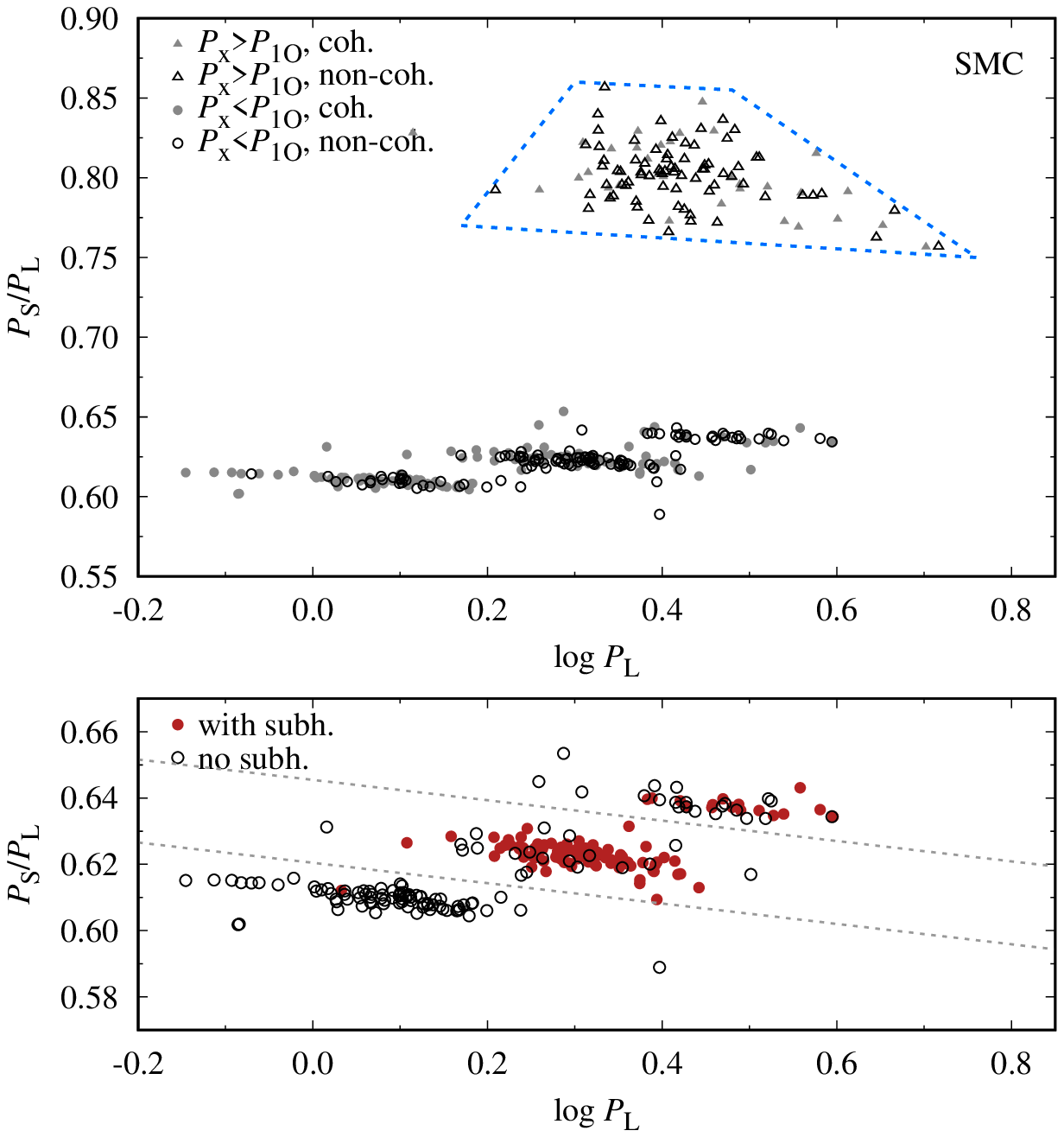}
\includegraphics[width=\columnwidth]{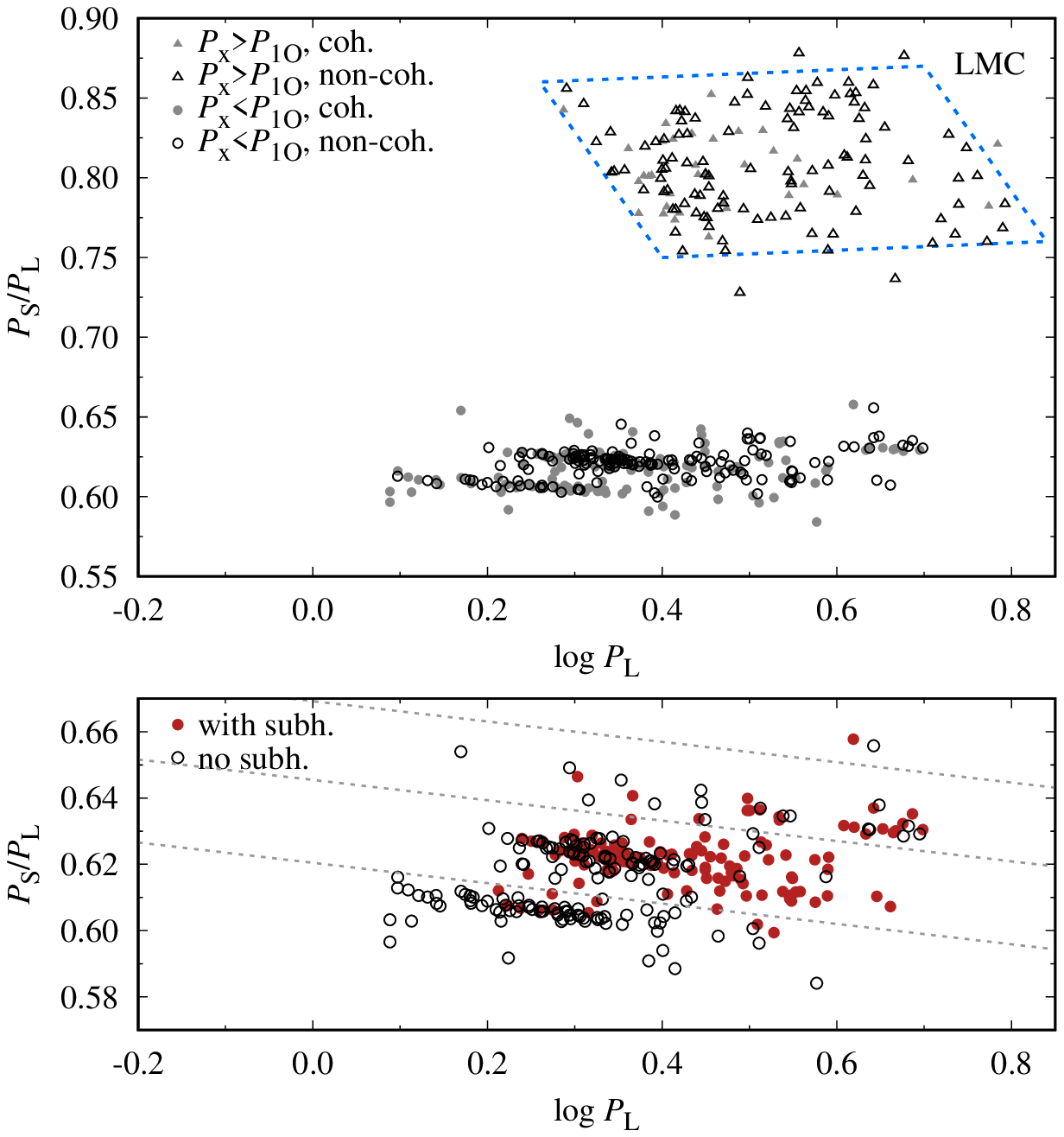}
\caption{Petersen diagrams for double-periodic stars with $\px/\pov\in(0.60,\,0.65)$ for the SMC (left panels) and the LMC (right panels). In the top panels, symbols are the same as in Fig.~\ref{fig:petall}. For stars for which power was also detected at around $1/2\fx$, we show both $\px/\pov$ (circles) and $\pov/\psh$ (triangles). In the bottom panels, we show only the $\px/\pov$ ratios. Filled symbols correspond to stars for which power was detected at around $1/2\fx$.}
\label{fig:pet61}
\end{figure*}

In Fig.~\ref{fig:pet61} we show the Petersen diagrams for the discussed group of stars, separately for the SMC (left) and for the LMC (right). The top panels show both the $\px/\pov$ period ratios (circles) and, for stars in which signals centred at $1/2\fx$ were detected, $\pov/\psh$ ratios (triangles), where $\psh=1/\fsh$ is a period corresponding to the highest peak within the power excess detected near $1/2\fx$. In the top panels, symbols are exactly the same as in Fig.~\ref{fig:petall}; in particular coherent signals are marked with filled symbols, while non-coherent signals are marked with open symbols. In the bottom panels, we zoom into the $\px/\pov$ period ratios only. This time, filled circles correspond to stars for which signals at $1/2\fx$ were detected, while for stars plotted with open symbols no signal at sub-harmonic frequency could be detected.

For the further analysis, particularly to test the model proposed to explain the nature of the discussed periodicities, it is crucial to divide the stars into sequences, based on their location in the Petersen diagram. Division solely based on $\px/\pov$ is not possible, as sequences are slanted. While in the case of the SMC nearly all stars can be assigned to a particular sequence without any ambiguity, in the LMC it is not possible; hence our procedure is based on the SMC stars. First, we divided the SMC stars into three sequences manually, in full agreement with the division lines plotted with the dashed lines in the bottom left panel of Fig.~\ref{fig:pet61} (lines were computed later on). Then, we fitted a linear function ($\px/\pov$ as a function of $\log\pov$) separately for the three sequences; results are collected in the top section of Tab.~\ref{tab:slopes}. Slopes for the three sequences are similar. Finally, we projected the location of all stars in the Petersen diagram on a line perpendicular to the reference line, for which we choose the linear fit to the middle sequence, crossing through an arbitrary reference point, at $\log\pov=0.3$ (the exact value has no impact on further analysis). 

\begin{figure}
\includegraphics[width=\columnwidth]{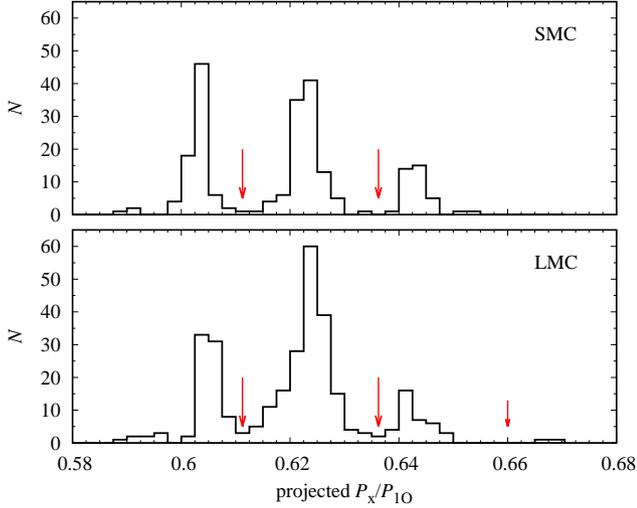}
\caption{Distributions of the projected period ratios, $\px/\pov$, for the SMC (top panel) and the LMC (bottom panel).}
\label{fig:histseq}
\end{figure}

In Fig.~\ref{fig:histseq}, we show the resulting distribution of the projected period ratios in the SMC and in the LMC. The existence of three sequences is clear also in the LMC. We can select two values of the projected period ratios, marked with long arrows in Fig.~\ref{fig:histseq}, the same for the LMC and the SMC, that coincide with the minima of the distributions. These are used to separate the stars into three sequences. It is also clear that two stars in the LMC with the largest projected period ratios should not be included in the top sequence, but rather form a {\it fourth} sequence. The boundary value is arbitrarily put at the projected period ratio of 0.66 and marked with a short arrow in the bottom panel of Fig.~\ref{fig:histseq}. The corresponding division lines are plotted with dashed gray lines in the Petersen diagrams in Fig.~\ref{fig:pet61}. For the SMC, the resulting three sequences fully agree with the prior manual division. 

We can now fit the linear functions into three sequences in the LMC (except for the fourth sequence with two stars only); results are collected in the bottom part of Tab.~\ref{tab:slopes}. At this point, we conclude that the slopes are all very similar; for the middle sequence they are the same, within 1$\sigma$, for the LMC and the SMC. Thus we consider our division into the sequences as final, keeping in mind that for the LMC, assignment of a few stars that fall in between the sequences is necessarily tentative.

\begin{table}
\caption{Coefficients of the linear fits, $\px/\pov=a\cdot\log\pov+b$, to the three sequences in the SMC (top section) and in the LMC (bottom section).}
\label{tab:slopes}
\begin{tabular}{lrr}
\hline
Sequence    & Slope, $a$          & Intercept, $b$\\
\hline
SMC, bottom & $-0.0265\pm 0.0040$ & $0.6116\pm 0.0005$\\
SMC, middle & $-0.0307\pm 0.0042$ & $0.6319\pm 0.0013$\\
SMC, top    & $-0.0331\pm 0.0059$ & $0.6537\pm 0.0027$\\
\hline
LMC, bottom & $-0.0318\pm 0.0043$ & $0.6138\pm 0.0013$\\
LMC, middle & $-0.0319\pm 0.0033$ & $0.6332\pm 0.0013$\\
LMC, top    & $-0.0378\pm 0.0036$ & $0.6560\pm 0.0020$\\
\hline
\end{tabular}
\end{table}

The number of stars within each of the three sequences is given in Tab.~\ref{tab:seqprop}. Within each sequence, we have counted the stars with non-coherent signal at $\fx$ and with signals detected close to $1/2\fx$. Numbers and incidence rates (within each sequence) are also collected in Tab.~\ref{tab:seqprop}.

\begin{table}
\caption{Properties of the three sequences in the SMC and LMC: numbers of stars within each of the sequences, $N$, including stars with non-coherent signals at $\fx$ and stars with signals detected close to $1/2\fx$. Incidence rates for the last two groups are also given.}
\label{tab:seqprop}
\begin{tabular}{lrrrr}
\hline
        &    SMC   &                & LMC    &  \\
        &   $N$ & \% & $N$ & \% \\
\hline
bottom seq. &   79 &              &   84 &               \\
non-coh.     &   25 & $31.7\pm5.2$ &   27 & $32.1\pm5.1$  \\
with sh     &    1 &  $1.3\pm1.3$ &   10 & $11.9\pm3.5$  \\
\hline
middle seq.  &  106 &              &  184 &               \\
non-coh.      &   58 & $54.7\pm4.8$ &  108 & $58.7\pm3.6$  \\
with sh      &   86 & $81.1\pm3.8$ &  113 & $61.4\pm3.6$  \\
\hline
top seq.    &   37 &              &   36 &               \\
non-coh.     &   26 & $70.3\pm7.5$ &   18 & $50.0\pm8.3$  \\
with sh     &   16 & $43.2\pm8.1$ &   19 & $52.8\pm8.3$  \\
\hline
\end{tabular}
\end{table}

Based on Figs~\ref{fig:pet61} and \ref{fig:histseq}, and Tabs~\ref{tab:slopes}, and \ref{tab:seqprop}, we observe the following.

(i) The $\px/\pov$ period ratios form three well populated sequences in the Petersen diagram, both for SMC and LMC. The sequences are slanted; in the SMC we observe a trend of increasing slope as we move from the bottom to the top sequence. In the LMC the slopes for the bottom and the middle sequence are the same, within errors, and the largest slope is for the top sequence (Tab.~\ref{tab:slopes}). In the LMC, a much larger dispersion of period ratios, as compared to the SMC is present. The three sequences are not equally populated (Tab.~\ref{tab:seqprop}). In both Clouds, middle sequence is the most populated and the top sequence is the least populated.

(ii) In 15 stars in the LMC and in 3 stars in the SMC we simultaneously detect signals corresponding to two sequences (two rows are given in Tab.~\ref{tab:61}). Assuming the \cite{wd16} model (see Sect.~\ref{ssec:discussion61}) correctly tights these signals to excitation on moderate-degree non-radial modes, such stars are interesting targets for asteroseismic modelling \citep[see][for application to RR~Lyr stars with analogous form of variability]{NS22}. The same applies to double-mode radial Cepheids with additional signals -- a few candidate stars were reported in the preceding sections.

(iii) In the LMC we tentatively identified a fourth sequence with the largest period ratios. It consists of only 2 stars. The presence of power excess centred at $1/2\fx$ in one of them indicates that we observe the same form of pulsation as for the other three sequences. 

(iv) In both SMC and LMC we also observe stars with low period ratios, located below the main progression of the bottom sequences. In Fig.~\ref{fig:histseq} these stars, with projected period ratios below $0.5975$, may be considered either as a short period ratio tail of the distribution for the bottom sequence, or as a separate sequence. We adopt the former option, and include these stars in the bottom sequence. 

(v) Non-coherent signals at $\fx$ are common (Tab.~\ref{tab:seqprop}). In the SMC they are most common for the top sequence ($70.3\pm7.5$ per cent), then for the middle sequence ($54.7\pm4.8$ per cent) and finally for the bottom sequence ($31.7\pm5.2$ per cent). In the LMC, the numbers are similar, except for the top sequence for which incidence rate is lower ($50.0\pm8.3$ per cent).

(vi) Since $\px/\pov$ period ratios form three sequences in the Petersen diagram, one would expect that period ratios for sub-harmonic signals, $\pov/\psh$, should also fall along three sequences. As is visible in the top panels of Fig.~\ref{fig:pet61}, this is not the case. This is because majority of signals detected at around $1/2\fx$ ($69.8\pm4.5$ per cent in the SMC and $76.2\pm3.5$ per cent in the LMC) are non-coherent, and appear in the frequency spectrum as a band of increased power, which we characterise through the highest peak within. Its frequency, $\fsh$, may be off $1/2\fx$. The point was already illustrated in Fig.~\ref{fig:fspexample} for a few stars: signals are well centred at $1/2\fx$ (marked with arrow), however the highest peak (marked with a filled circle) is off. Still, each signal detected at around $1/2f\fx$ is assigned to a specific sequence as marked in the bottom panels of Fig.~\ref{fig:pet61} (filled symbols). 

\begin{figure}
\includegraphics[width=\columnwidth]{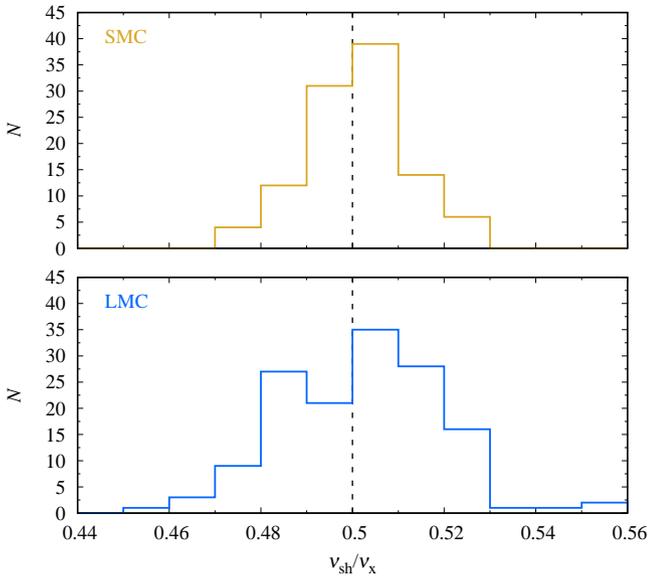}
\caption{Distributions of the $\fsh/\fx$ frequency ratios for the SMC (top panel) and the LMC (bottom panel).}
\label{fig:shdist}
\end{figure}

In Fig.~\ref{fig:shdist} we plot the distribution of $\fsh/\fx$ ratios for the SMC and LMC stars. The distributions are centred on $0.5$, but with wide and asymmetric wings. For $55.7$ per cent of the SMC and $57.1$ per cent of the LMC stars we have $\fsh/\fx>0.5$. It indicates that more often $\fsh>0.5\fx$ for the highest peak within power excess around $1/2\fx$. This is the case for all frequency spectra illustrated in Fig.~\ref{fig:fspexample}.

(vii) As quantified in Tab.~\ref{tab:seqprop}, signals around $1/2\fx$ are most commonly detected for the middle sequence ($81.1\pm3.8$ and $61.4\pm3.6$ per cent for the SMC and the LMC, respectively) and then for the top sequence ($43.2\pm8.1$ and $52.8\pm8.3$ per cent for the SMC and the LMC, respectively). For the bottom sequence only a single detection was reported in the SMC ($1.3\pm1.3$ per cent), but significantly more in the LMC (10 detections, $11.9\pm3.5$ per cent). In the tentatively identified fourth sequence in the LMC, one out of two stars shows signal at $1/2\fx$.

\begin{figure}
\includegraphics[width=\columnwidth]{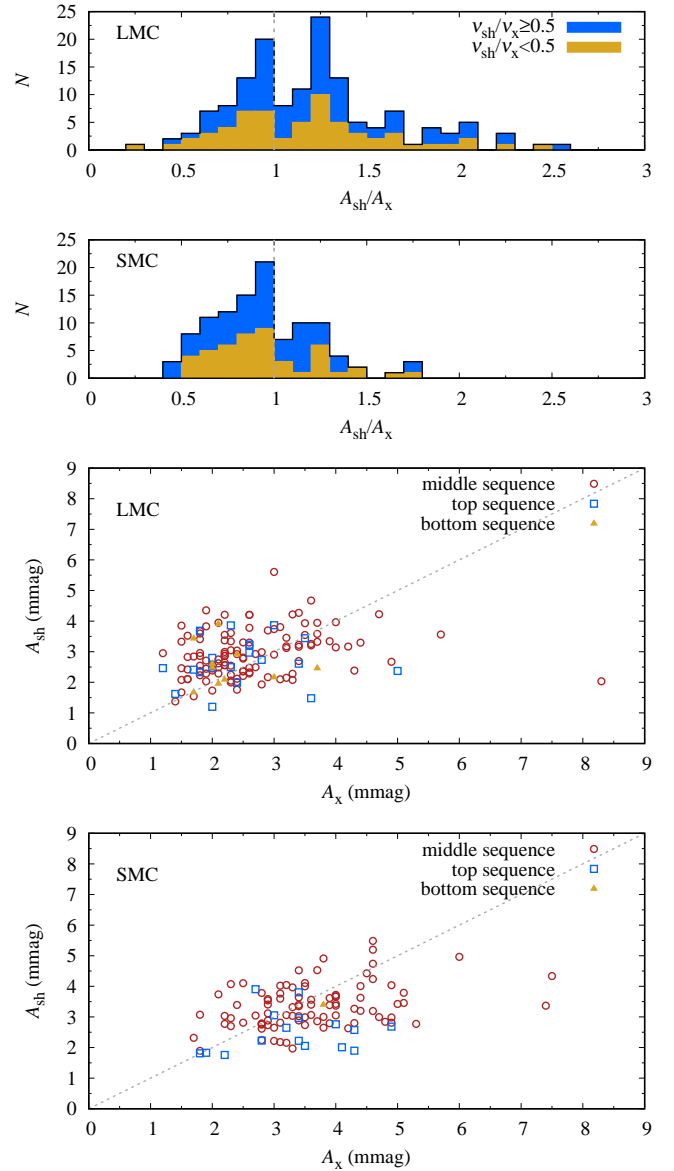}
\caption{Distribution of amplitude ratios, $\ash/\ax$ (top panels) and plots of
$\ash$ versus $\ax$ (bottom panels). In the top panel, contributions from signals with $\fsh/\fx$ larger and smaller than 0.5 are indicated with different colors, as given in the legend. In the bottom panel, signals corresponding to different sequences are plotted with different symbols.}
\label{fig:61shamps}
\end{figure}

(viii) The distribution of relative amplitudes of signals centred at $1/2\fx$ with respect to amplitudes of signals at $\fx$, $\ash/\ax$, is asymmetric, as illustrated in the top panels of Fig.~\ref{fig:61shamps}. The coloured bars show relative contributions of signals with $\fsh/\fx$ larger and smaller than 0.5. We observe that in the LMC $\ash>\ax$ is more common (63\, per cent) than $\ash<\ax$, while reverse is true for the SMC, as $\ash>\ax$ holds for 35\, per cent of signals.

(ix) The distribution of amplitudes of signals corresponding to specific sequences in the Petersen diagram follows the distribution outlined above -- see bottom panels of Fig.~\ref{fig:61shamps} in which $\ash$ is plotted against $\ax$, separately for both Clouds and three sequences. In the LMC we observe that typically $\ash>\ax$ for signals corresponding to the top and middle sequences. In the SMC we observe that typically $\ash<\ax$ for all sequences; in particular for nearly all signals of the top and bottom (1 signal) sequences.

\subsection{Period ratios $\pov/\px$ around 0.684}\label{ssec:686}

\begin{figure}
\includegraphics[width=\columnwidth]{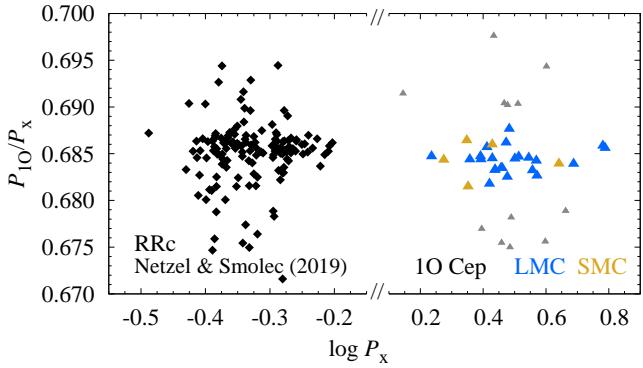}
\caption{The Petersen diagram zoomed around double-periodic stars clustered at $\pov/\px\approx0.684$. A group of double-periodic RRc stars of similar characteristics are included for a comparison.}
\label{fig:pet686}
\end{figure}

\begin{figure}
\includegraphics[width=\columnwidth]{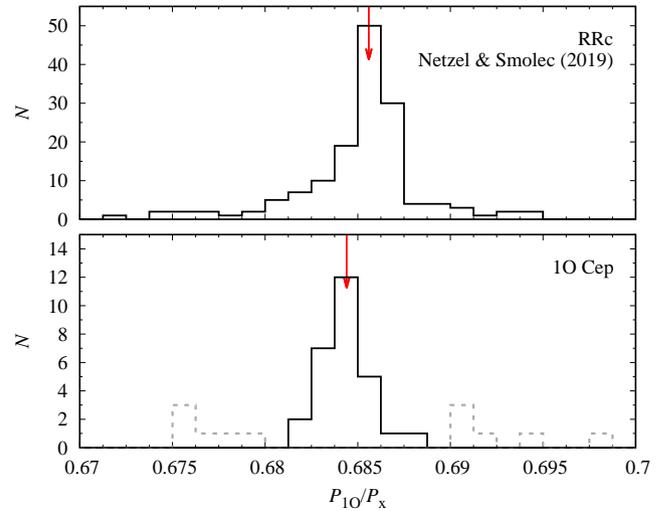}
\caption{Distribution of period ratios for double-periodic stars with dominant 1O and additional coherent periodicity with $\pov/\px$ in the $0.67-0.70$ range. Top panel shows RRc stars and bottom panel shows 1O Cepheids.}
\label{fig:hist686}
\end{figure}

In the Petersen diagram in Fig.~\ref{fig:petall}, an over-density of double-periodic pulsators at $\log\plong\approx 0.5$ and $\pshort/\plong\approx 0.684$ is well visible in the LMC. In all these stars $\px>\pov$ and the detected additional signal is coherent -- stars are plotted with filled triangles. A zoom into the relevant part of the Petersen diagram, showing both SMC and LMC stars, is shown in the right side of Fig.~\ref{fig:pet686}. The number of double-periodic pulsators in the SMC is much smaller; in the following stars form both Clouds are analysed together. An over-density is well captured in the histogram of period ratios for double-periodic stars with $\pov/\px$ in the $0.67-0.70$ range displayed in the bottom panel of Fig.~\ref{fig:hist686}. 

The discussed double-periodic Cepheids share the properties of double-periodic RRc stars discovered by \cite{nsd15} and studied in more detail by \cite{ns19}. The Petersen diagram for double-periodic RRc stars from the latter study is displayed in the left side of Fig.~\ref{fig:pet686} and the corresponding histogram of period ratios is displayed in the top panel of Fig.~\ref{fig:hist686}. In RRc stars, the distribution is peaked around the median value, $0.6856$ (marked with red arrow), with wide wings (period ratios in between $0.67$ and $0.70$). For first overtone Cepheids, we define an equivalent group of double-periodic stars, with period ratios in between $0.68$ and $0.69$. Five stars belong to SMC ($0.3\pm0.1$\,per cent) and 23 to LMC ($1.3\pm0.3$\,per cent). These stars are marked with large colored triangles in Fig.~\ref{fig:pet686} and the corresponding histogram is plotted with black solid line in Fig.~\ref{fig:hist686}. The distribution peaks around the median value, $0.6844$ (marked with red arrow) -- a bit lower than for the RRc stars. Data for these stars are collected in Tab.~\ref{tab:1o686}. The detections are firm, with $\sn$ usually exceeding $5.0$. Except for \idlmc{2137}, in which significant peaks at $\fov+\fx$ and $\fov-\fx$ are detected, we find no combinations with the radial 1O mode.

For RRc stars, the amplitude of the additional periodicity amounts to $3.3$ per cent of the 1O amplitude, on average. For 1O Cepheids the number is similar, amplitude of the additional signals is $3.9$ per cent of the 1O amplitude, on average.

In Fig.~\ref{fig:pet686} there are more double-periodic Cepheids with additional coherent signals and period ratios in the $0.67-0.70$ range (marked with small grey triangles in the right part of Fig.~\ref{fig:pet686}). The corresponding distribution of period ratios is plotted with grey dashed line in Fig.~\ref{fig:hist686}. Whether these are double-periodic stars of the same type, or not, is an open question.

\begin{table*}
\caption{Candidates for double periodic pulsators with dominant radial first overtone mode and additional periodicity with period ratios clustered around $\pov/\px=0.684$. Consecutive columns contain: star's id, first overtone period, $\pov$, additional, longer period, $\px$ , corresponding period ratio, $\pov/\px$, amplitude of the first overtone, $\aov$, amplitude of the additional periodicity, $\ax$, signal-to-noise for the detection of additional periodicity and remarks: `al' -- daily alias of a signal at $\fx$ is higher, `nsO' -- non-stationary first overtone, `cf' -- combination frequency of $\fx$ and $\fov$ detected, `ap'-- additional periodicity detected, `0.61' -- additional periodicity with $\px/\pov\in(0.60,\,0.65)$ was detected, `tdp' -- time-dependent analysis was conducted.}
\label{tab:1o686}
\begin{tabular}{lrrrrrrr}
\hline
Star & $\pov$\,(d) & $\px$\,(d) &  $\pov/\px$ & $\aov$ (mag) & $\ax$ (mag) & $\sn$    & Remarks\\
\hline
\idlmc{0113} & 1.838337(4) & 2.6856(3) & 0.6845 & 0.0780(3) & 0.0019(3) &  4.6 &  \\
\idlmc{0184} & 1.793580(5) & 2.6308(3) & 0.6818 & 0.0708(3) & 0.0026(3) &  5.8 &  \\
\idlmc{0193} & 1.691671(4) & 2.4717(2) & 0.6844 & 0.0961(5) & 0.0038(5) &  6.5 & ap \\
\idlmc{0676} & 2.544534(7) & 3.7275(4) & 0.6826 & 0.0887(4) & 0.0035(4) &  6.6 & ap \\ 
\idlmc{0817} & 1.873575(3) & 2.7420(4) & 0.6833 & 0.1210(3) & 0.0019(3) &  4.7 & ap\\
\idlmc{0925} & 2.045508(6) & 2.9970(4) & 0.6825 & 0.0786(4) & 0.0021(4) &  4.6 & ap\\
\idlmc{1346} & 2.462782(9) & 3.6047(4) & 0.6832 & 0.1101(6) & 0.0029(3) &  6.5 & 0.61, F1O, ap, nsO, tdp\\
\idlmc{1801} & 1.178183(2) & 1.7207(1) & 0.6847 & 0.0812(4) & 0.0025(4) &  5.0 & \\
\idlmc{2137} & 4.21030(2)  & 6.1411(4) & 0.6856 & 0.0785(2) & 0.0068(2) & 16.5 & cf, F1O \\
\idlmc{2172} & 1.955678(4) & 2.8615(3) & 0.6834 & 0.0778(3) & 0.0023(3) &  5.9 & nsO\\
\idlmc{2406} & 2.091337(4) & 3.0413(4) & 0.6876 & 0.0926(3) & 0.0023(3) &  5.3 & ap, F1O\\
\idlmc{2432} & 1.686911(2) & 2.4637(3) & 0.6847 & 0.1118(3) & 0.0023(3) &  5.3 & \\
\idlmc{2444} & 2.232034(4) & 3.2600(3) & 0.6847 & 0.1079(3) & 0.0031(3) &  7.5 & nsO\\
\idlmc{2597} & 1.972834(4) & 2.8865(3) & 0.6835 & 0.0909(3) & 0.0022(3) &  5.3 & 0.61\\
\idlmc{2789} & 2.537724(9) & 3.7088(3) & 0.6842 & 0.0665(3) & 0.0047(3) &  8.9 & \\
\idlmc{2895} & 2.170233(5) & 3.1706(3) & 0.6845 & 0.0730(4) & 0.0030(4) &  5.5 & \\
\idlmc{3033} & 1.770260(3) & 2.5817(3) & 0.6857 & 0.0958(4) & 0.0021(4) &  4.3 & ap \\
\idlmc{3674} & 2.039138(6) & 2.9717(4) & 0.6862 & 0.0785(5) & 0.0030(5) &  4.9 & \\
\idlmc{3827} & 1.870824(4) & 2.7381(2) & 0.6833 & 0.0886(4) & 0.0033(3) &  6.5 & nsO\\
\idlmc{4494} & 4.14287(2)  & 6.0405(9) & 0.6858 & 0.0792(3) & 0.0032(3) &  6.7 & nsO\\
\idlmc{4503} & 3.33323(2)  & 4.8739(5) & 0.6839 & 0.0668(4) & 0.0046(4) &  7.8 & ap, nsO\\ 
\idlmc{4505} & 2.401212(6) & 3.5077(4) & 0.6846 & 0.0972(4) & 0.0029(4) &  5.3 & ap\\
\idlmc{4624} & 1.559273(5) & 2.2784(2) & 0.6844 & 0.0745(6) & 0.0043(6) &  5.3 & F1O\\
\hline
\idsmc{0151} & 1.532852(3) & 2.2493(1) & 0.6815 & 0.1121(5) & 0.0048(5) &  6.9 & \\
\idsmc{1598} & 2.99418(2)  & 4.3778(9) & 0.6839 & 0.0653(5) & 0.0020(4) &  4.3 & nsO, tdp, al, ap\\
\idsmc{1634} & 1.288075(2) & 1.8822(2) & 0.6843 & 0.0980(6) & 0.0033(6) &  4.1 & \\
\idsmc{3158} & 1.528997(2) & 2.2275(1) & 0.6864 & 0.1348(5) & 0.0051(5) &  7.5 & \\
\idsmc{4909} & 1.84246(1)  & 2.6858(3) & 0.6860 & 0.0560(7) & 0.0041(7) &  4.5 & \\
\hline
\end{tabular}
\end{table*}


\subsection{Unclassified period ratios}\label{ssec:unclassified}

\begin{figure*}
\includegraphics[width=\columnwidth]{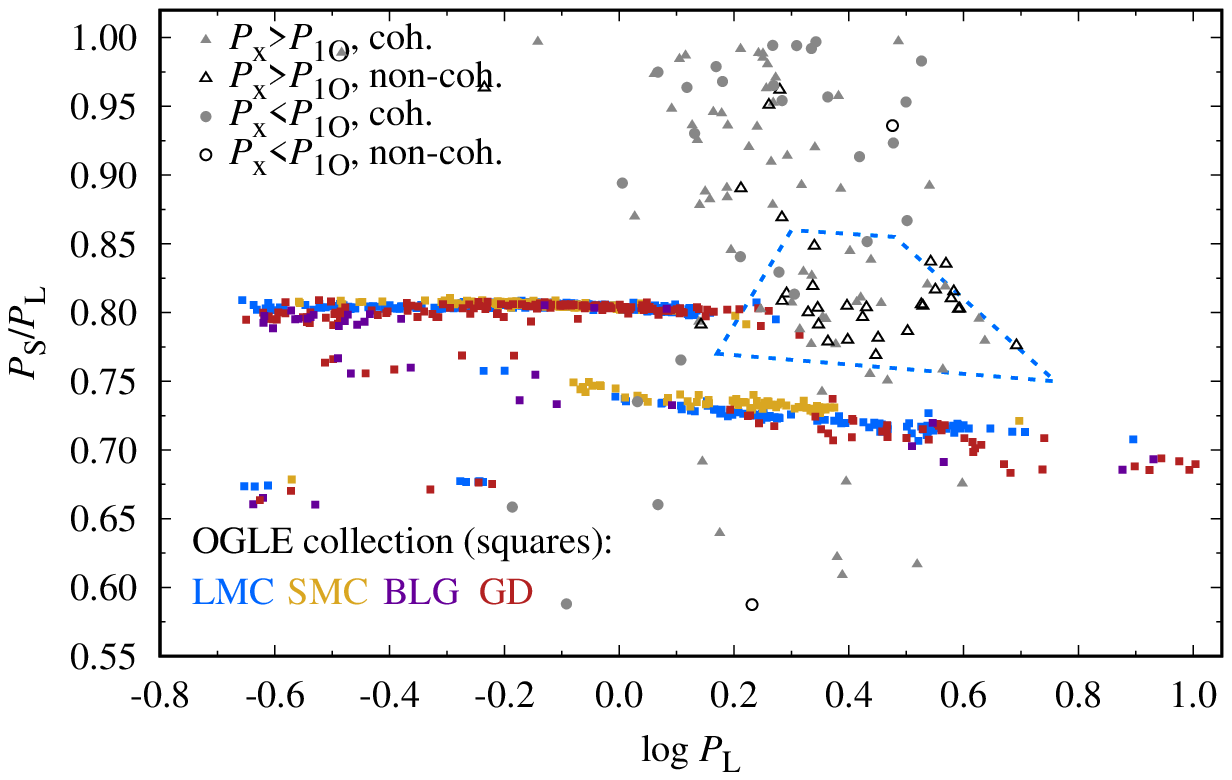}
\includegraphics[width=\columnwidth]{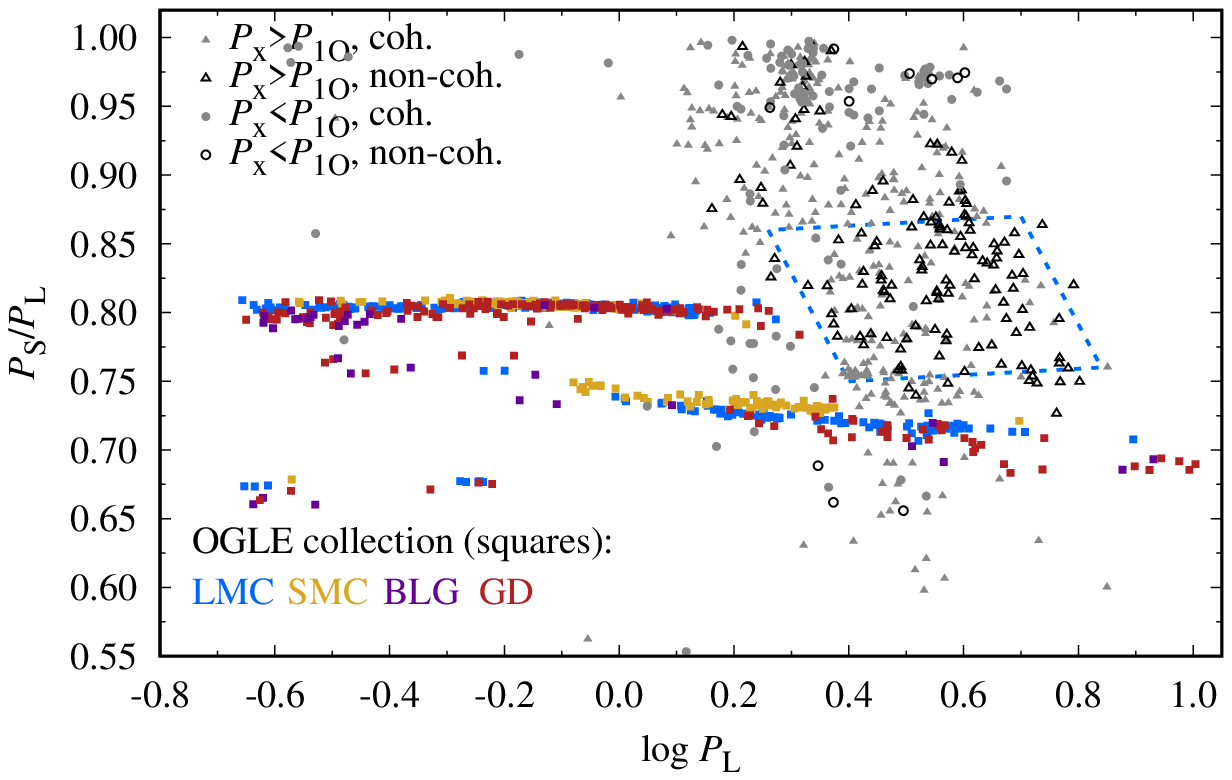}
\caption{The Petersen diagram with double-periodic pulsators after filtering out the new radial double-mode candidates, stars with $\px/\pov\in(0.60,\,0.65)$ and stars with $\pov/\px$ centred at $0.684$. Double-mode radial pulsators from OGLE-CVS are still included for context. The dashed polygons indicate the range at which sub-harmonic periodicities were detected in $\px/\pov\in(0.60,\,0.65)$ stars, as defined in Fig.~\ref{fig:pet61}.}
\label{fig:petadd}
\end{figure*}

Except signals that correspond to double-periodic pulsations described in the previous sections, other signals were detected in 153 1O Cepheids in the SMC ($8.6\pm0.7$\,per cent) and in 482 1O Cepheids in the LMC ($27.3\pm1.1$\,per cent). In many stars we detect more than one additional signal. In total, we have detected 168 additional signals in the SMC and 629 signals in the LMC. Period ratios for all these signals are plotted in Fig.~\ref{fig:petadd}, which is essentially Fig.~\ref{fig:petall} without candidates for double-mode radial pulsation, without double-periodic stars with $\px/\pov$ in the $0.60-0.65$ range, and without stars with period ratios centred at $\pov/\px\approx 0.684$. Basic properties of these stars are collected in Tab.~A3 in the on-line Appendix, section of which is shown in Tab.~\ref{tab:add} for a reference.

\begin{table*}
\caption{Properties of first overtone Cepheids with additional variability. The consecutive columns contain: star's id, first overtone frequency, $\fov$, frequency of the additional variability, $\fx$ , corresponding period ratio (shorter to longer), amplitude of the first overtone, $\aov$, and amplitude ratio, $\ax/\aov$, and remarks: `al' -- daily alias of signal at $\fx$ is higher; `nsx' -- complex appearance of the signal at $\fx$; `nsO' -- non-stationary first overtone; `cf' -- combination frequency of $\fx$ and $\fov$ detected; `tdp' -- time-dependent analysis was conducted.}
\label{tab:add}
\begin{tabular}{lrrrrrrl}
\hline
Star & $\fov$ (d$^{-1}$) & $\fx$ (d$^{-1}$) & $P_{\rm S}/P_{\rm L}$ & $\aov$ (mag) & $\ax/\aov$ & $\sn$ & Remarks\\
\hline
OGLE-SMC-CEP-0021 & 0.916019(1)  & 0.72656(5)  & 0.7932 & 0.2030(9) & 0.025 &  4.1 &  \\ 
OGLE-SMC-CEP-0065 & 0.575606(1)  & 0.55434(4)  & 0.9631 & 0.1514(7) & 0.028 &  4.8 &   \\ 
OGLE-SMC-CEP-0114 & 0.568231(2)  & 0.56159(2)  & 0.9883 & 0.0988(6) & 0.064 &  7.0 &  \\ 
OGLE-SMC-CEP-0125 & 0.540067(2)  & 0.54312(2)  & 0.9944 & 0.0861(6) & 0.086 &  8.5 &   \\ 
*                 & 0.540067(2)  & 0.41954(4)  & 0.7768 & 0.0861(6) & 0.038 &  5.0 &   \\ 
\ldots &  & & & & & & \\
\hline
\end{tabular}
\end{table*}

\begin{figure}
\includegraphics[width=\columnwidth]{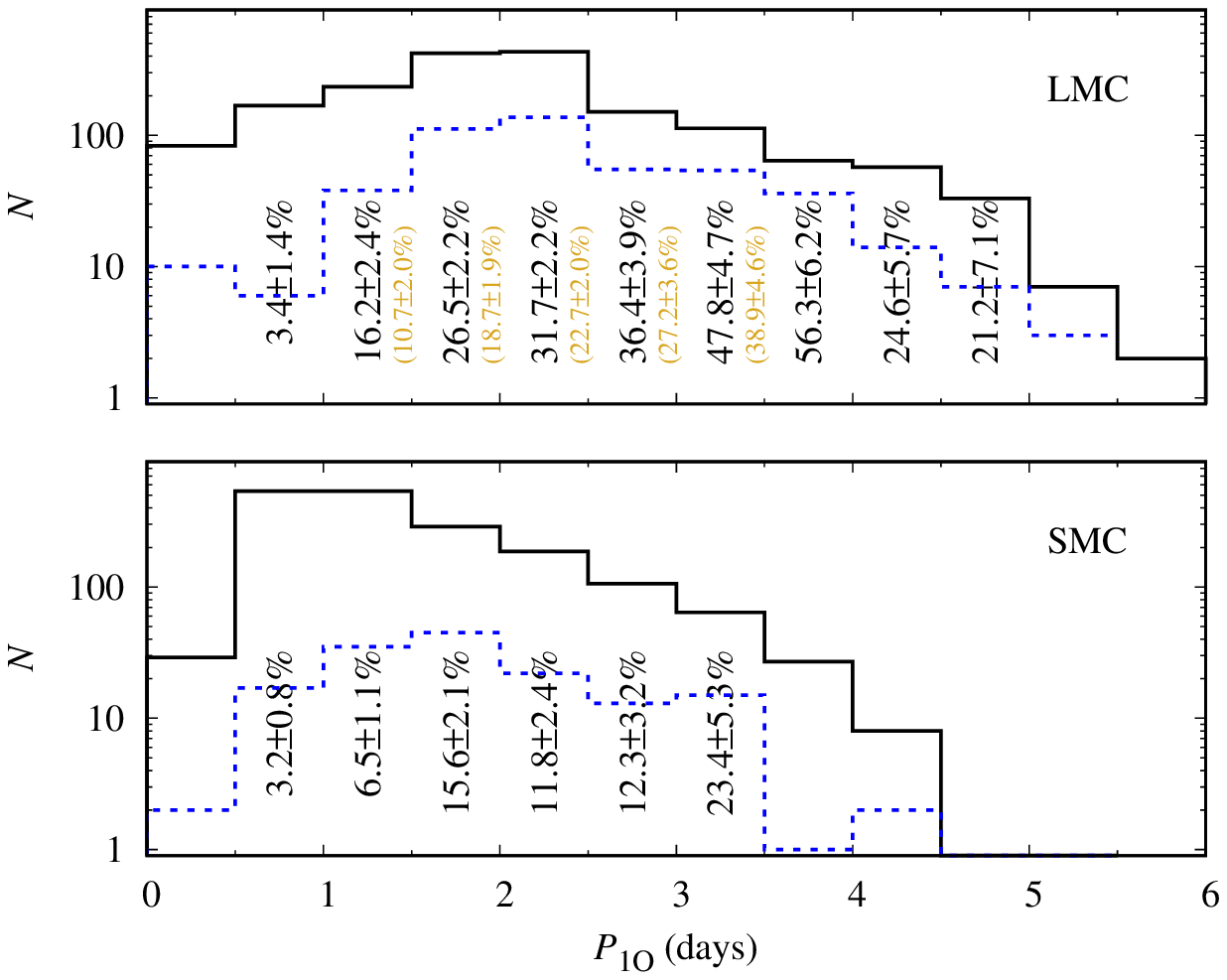}
\caption{First overtone period distributions for all Cepheids (solid black line) and for Cepheids with additional periodicities in the frequency spectrum (dashed blue line) in the LMC (top panel) and the SMC (bottom panel). For period bins with the largest number of stars incidence rates are given. For the LMC, the incidence rates given in parenthesis result from applying the SMC's detection limit.}
\label{fig:adddist}
\end{figure}

A striking difference between SMC and LMC, well visible in Fig.~\ref{fig:petadd}, and corroborated by incidence rates given at the beginning of this section, is many more additional signals detected in the latter system. In Fig.~\ref{fig:adddist} we show the distribution of first overtone periods for all 1O Cepheids and for those with additional signals detected, both for the LMC (top panel) and the SMC (bottom panel). Incidence rates are given for specific period bins with significant number of stars. In each period bin incidence rate is larger in the LMC. While the highest incidence rate in the SMC is $23.4$\,per cent (for $3.0<P_{\rm 1O}\leq 3.5$\,d), in the LMC it is $56.3$\,per cent (for $3.5<P_{\rm 1O}\leq 4.0$\,d). We stress that for first overtone periods between $3$\,d and $4$\,d $\sim$half of the 1O Cepheids in the LMC show additional signals in the frequency spectrum (and this assessment does not include the classes of signals discussed in the preceding sections).

The obvious explanation for significantly lower number of signals in the SMC may be higher detection limit, as the SMC Cepheids  \citep[$\mu_{\rm SMC}=18.977$\,mag,][]{SMCdist} are further away than the LMC Cepheids \citep[$\mu_{\rm LMC}=18.477$\,mag,][]{LMCdist}. This is not the case however, which we show by applying the (period-dependent) SMC's detection limit to the LMC signals.

 To estimate the detection limit in both Magellanic Clouds we follow the method used in \cite{ss16} and \cite{s17}. For every 1O Cepheid in a given Magellanic Cloud we fit an eight order Fourier series, remove severe (6-$\sigma$) outliers and conduct time-dependent prewhitening on a season-to-season basis. Fourier transform is then computed in the $(0,\,6\fov)$ range and the average noise level is computed. Data on spectra in which no strong signals remain ($\sn<5$) are used to construct four-times noise (detection limit) against pulsation period plots. Finally, low order polynomials are fitted to the date to yield the detection limit.

\begin{figure}
\includegraphics[width=\columnwidth]{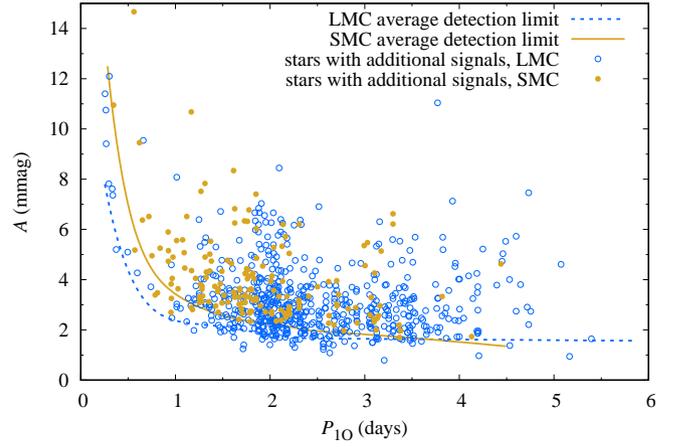}
\caption{Amplitude of additional signals plotted vs. $\pov$ for the LMC (open circles) and SMC (filled circles). Average detection limits are plotted with dashed and solid lines for the LMC and SMC, respectively.}
\label{fig:noise}
\end{figure}

The resulting detection limits in the SMC and LMC are plotted with solid and dashed curves in Fig.~\ref{fig:noise}, respectively, together with amplitudes of additional signals detected. For $\pov<4.0$\,d, detection limit is clearly higher in the SMC. We can now apply the SMC detection limit to LMC stars by removing from the sample those stars, for which amplitudes of additional signals fall below SMC detection limit. As a result, we arrive at incidence rates given in parenthesis in Fig.~\ref{fig:adddist}, just below the original incidence rates. While the incidence rates are lower, they are still well above the incidence rates reported in the SMC in the corresponding period bins.

We conclude that significantly higher incidence rate of additional signals in the LMC, as compared to the SMC, is an intrinsic property of 1O Cepheids and does not result from different detection limits for the two galaxies (for further discussion see Sect.~\ref{ssec:propSLMC}).

In the discussion below we will often distinguish between coherent and non-coherent signals. For the former ones, after prewhitening with the additional frequency, there is no significant remnant in the frequency spectrum. The signal can be well modeled with a single sine function with constant amplitude and phase. For non-coherent signals, residual power is observed after the prewhitening, either as unresolved significant peak, or as a power excess (which is often clear even before the prewhitening). The two classes of signals seem to have different properties.

\begin{figure}
\includegraphics[width=\columnwidth]{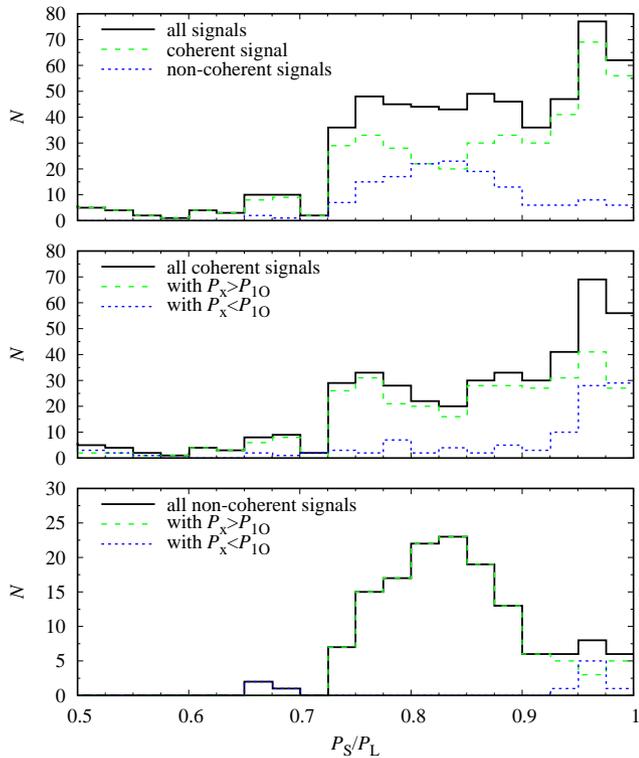}
\caption{Distribution of period ratios, shorter-to longer, $\pslr$, for 1O Cepheids with additional periodicities in the LMC. In the top panel contributions form stars with coherent and non-coherent signals are shown. In the middle and bottom panels, distributions for stars with coherent and non-coherent signals are shown separately, with contributions form stars with $\px<\pov$ and $\px>\pov$ plotted with different line styles.}
\label{fig:histadd}
\end{figure}

In the top panel of Fig.~\ref{fig:histadd}, we show the distribution of period ratios of additional signals with respect to first overtone period for the LMC. Note that we always plot shorter-to-longer period ratio, $\pslr$, thus either $\px/\pov$ or $\pov/\px$ is used. The distribution peaks for $\pslr>0.95$ (the fewer number of stars in the right-most bin, $\pslr>0.975$, as compared to the adjacent bin, is probably due to the fact that for the highest period ratios the signals may be unresolved with the first overtone frequency), then we observe a plateau down to $\pslr=0.75$ and decline toward lower period ratios. Green long-dash and blue short-dash lines show contributions from the coherent and non-coherent signals, respectively. In the middle and bottom panels, distributions for coherent and non-coherent signals are shown separately, this time also showing the contributions of signals for which $\px>\pov$ (green, long-dash line) and $\px<\pov$ (blue, short-dash line). We observe that the peak for high period ratios is nearly exclusively due to coherent signals. For lower period ratios, non-coherent signals appear and peak for $0.75<\pslr<0.9$. Together with coherent signals they build a plateau observed for $0.75<\pslr<0.95$. Below $\pslr=0.75$ the number of stars with both coherent and non-coherent additional periodicities gradually decreases.

It is also clear that for majority of signals, both coherent and non-coherent $\fx<\fov$ ($\px>\pov$). While coherent signals with $\px<\pov$ are still detected, there are only a few non-coherent signals with period shorter than $\pov$. 

Distributions and conclusions for the the SMC are qualitatively the same, with the exception that period ratios peak not only for high period ratios ($\pslr>0.95$), but there is a second peak within the plateau, at $0.80<\pslr<0.85$, resulting from the peak observed for non-coherent signals.

\begin{figure}
\includegraphics[width=\columnwidth]{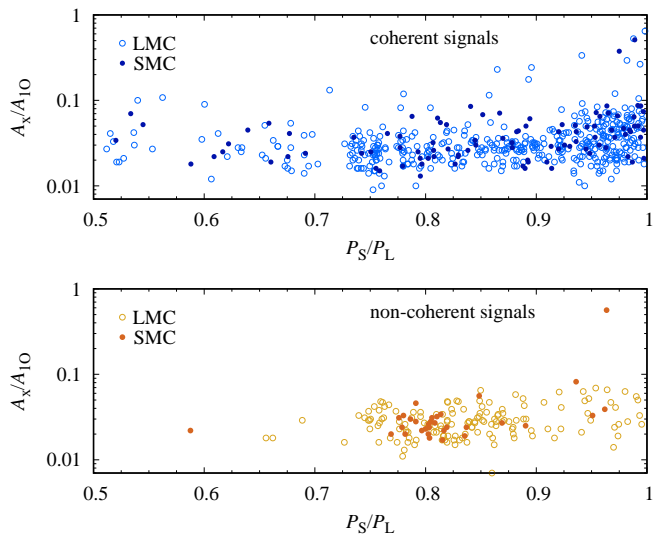}
\caption{Relative amplitude ratio, $\ax/\aov$, plotted against period ratio, $\pslr$ for 1O Cepheids with coherent additional signals in the frequency spectrum (top panel) and non-coherent ones (bottom panel). LMC and SMC stars are plotted with different symbols as indicated in the key. Note the logarithmic scale on y axis.}
\label{fig:ampadd}
\end{figure}

\begin{figure}
\includegraphics[width=\columnwidth]{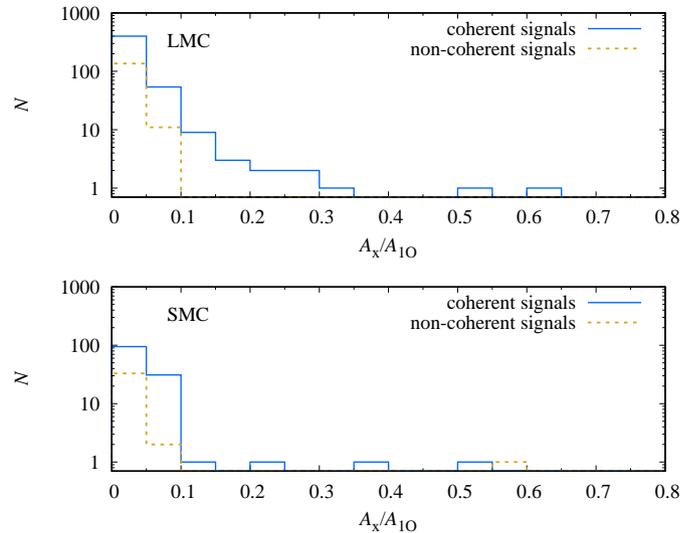}
\caption{Distribution of relative amplitudes, $\ax/\aov$, for 1O Cepheids with additional periodicities in the frequency spectrum in the LMC (top panel) and the SMC (bottom panel). Distributions for coherent and non-coherent signals are plotted with solid and dashed lines, respectively.}
\label{fig:ampadd_hist}
\end{figure}

In Fig.~\ref{fig:ampadd}, we show the relative amplitude ($\ax/\aov$) of additional signals plotted against the corresponding period ratio, $\pslr$, separately for the coherent signals (top panel) and non-coherent signals (second panel). LMC and SMC stars are plotted with different symbols. In Fig.~\ref{fig:ampadd_hist}, we plot the distributions of relative amplitudes for the LMC and SMC showing contributions of coherent and non-coherent signals. Large relative amplitudes (above $10$\,per cent) are observed exclusively for coherent signals (the only exception is relative amplitude of $56$\,per cent for one non-coherent signal in the star form SMC.
For nearly all non-coherent signals relative amplitude is below $8$\,per cent. For coherent signals, very high relative amplitudes are observed for high period ratios. Also, for coherent signals, we observe that average relative amplitude is higher for period ratios above $\sim0.95$.

The observations made so far suggest that we can distinguish two classes of additional periodicities. In the first class, additional periodicities cluster around first overtone (period ratios above $0.95$, both $\px>\pov$ -- more frequent and $\px<\pov$), are coherent and may have high relative amplitudes. As we discuss later in Sect.~\ref{ssec:cnm} they may originate from non-radial modes with frequencies close to the radial mode frequency. In the second class, additional periodicities have period ratios in between $0.75$ and $0.9$ ($\px>\pov$), are non-coherent and have typically low relative amplitudes. As we argue later in Sect.~\ref{ssec:789noharm}, these signals may originate form excitation of moderate degrees, $\ell=7,\,8,\,9$ modes, in fact the same class as already discussed in Sect.~\ref{ssec:61}, but for which harmonics, that then form three distinct sequences in the Petersen diagram, are not detected.




\subsection{Periodic Modulation of Pulsation}\label{ssec:modulation}

\begin{figure}
\noindent\includegraphics[width=\columnwidth]{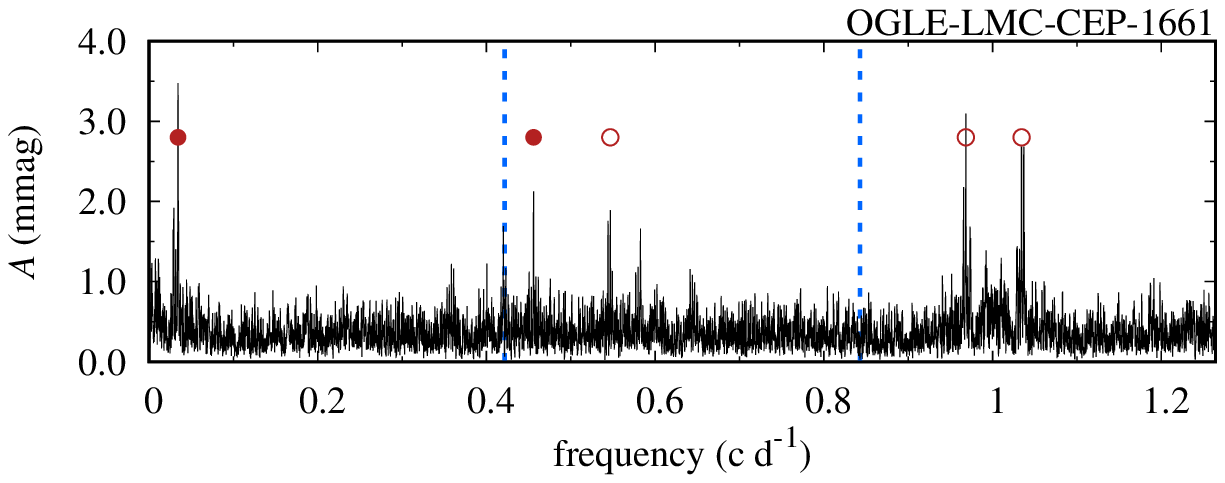}\\
\noindent\includegraphics[width=\columnwidth]{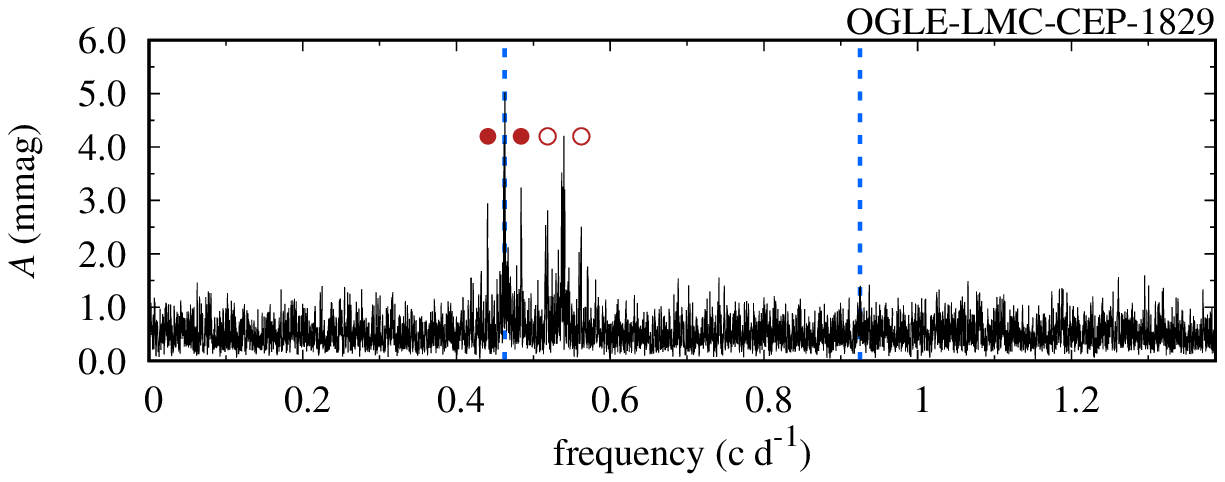}\\
\noindent\includegraphics[width=\columnwidth]{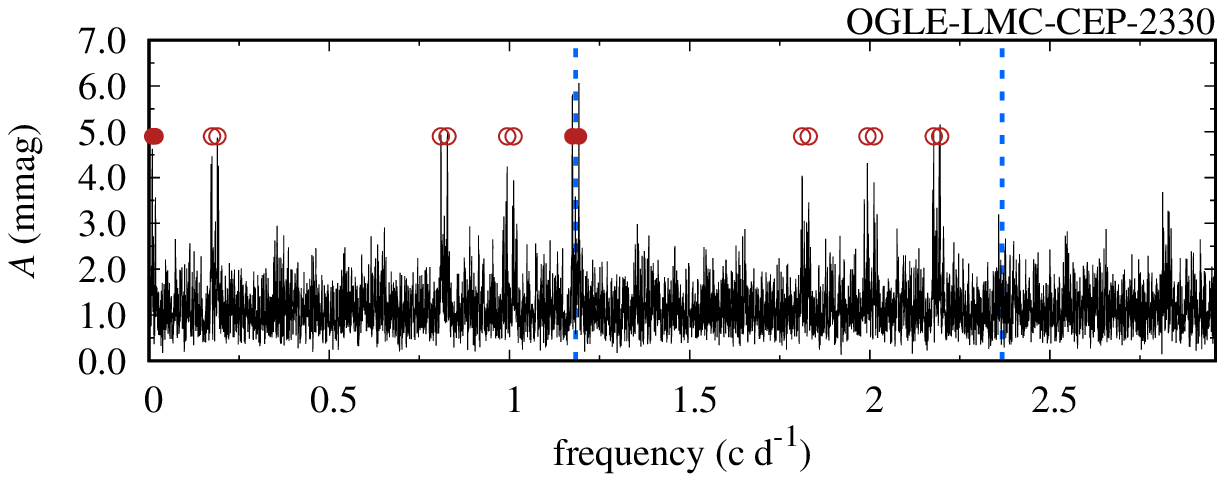}\\
\noindent\includegraphics[width=\columnwidth]{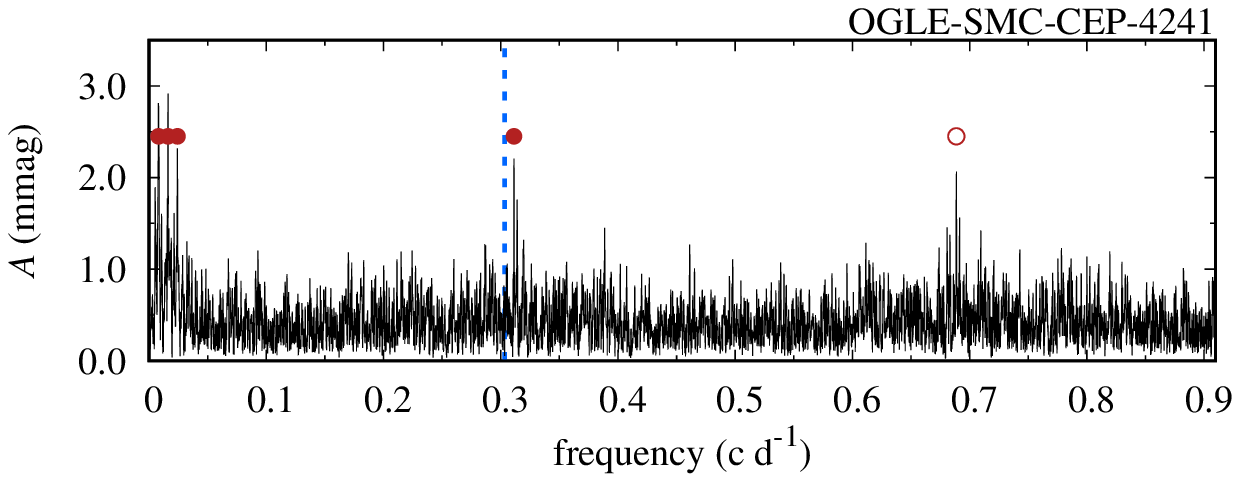}
\caption{Exemplary frequency spectra for modulated Cepheids, after prewhitening with the first overtone frequency and its harmonics (if present; dashed lines). Filled circles mark the modulation peaks, open symbols mark locations of the corresponding daily aliases.}
\label{fig:fspmodul}
\end{figure}

Periodic modulation of pulsation in single-mode Cepheids was rarely reported in the literature. \cite{o4_clouds_multimode} claimed long-period, large-amplitude modulation of pulsation in a few 1O Cepheids based on the appearance of light curves only. \cite{s17} detected a periodic modulation of pulsation in OGLE photometry for a sizeable sample of F-mode Magellanic Clouds Cepheids and a single modulated 1O Cepheid was reported by \cite{Kotysz}. Here we report a sample of 24 modulated 1O Cepheids in the LMC ($1.4\pm0.3$\, per cent) and only 3 in the SMC ($0.2\pm0.1$\, per cent). In each of these stars at least two modulation peaks were detected. Exemplary frequency spectra are illustrated in Fig.~\ref{fig:fspmodul}; see also remarks on individual stars later in this section. In Tab.~\ref{tab:mod} we provide basic properties of these stars: pulsation and modulation periods, amplitudes of the first overtone and most significant modulation peaks, full list of detected modulation peaks and remarks.

The SMC sample is scarce and two of the stars appear peculiar compared to the LMC sample (nonlinear modulation in \idsmc{4241} and very long modulation period in \idsmc{4372}). Consequently, the following discussion of modulation properties is based on the LMC sample only. 

\begin{figure}
\noindent\includegraphics[width=\columnwidth]{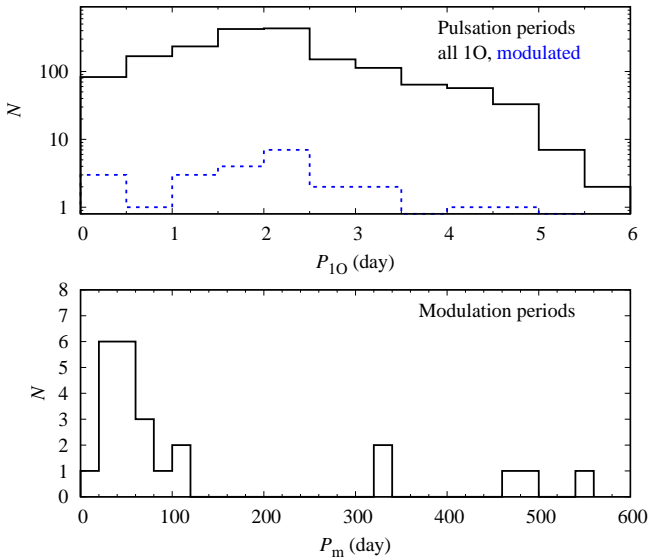}
\caption{Top panel: distribution of pulsation periods of all (black solid line) and modulated (blue dotted line) first overtone Cepheids in the LMC. Bottom panel: distribution of modulation periods in the LMC.}
\label{fig:hist_modulP}
\end{figure}

In the top panel of Fig.~\ref{fig:hist_modulP}, we show the distribution of pulsation periods of modulated stars (blue dashed-line), compared with overall period distribution of 1O Cepheids in the LMC (black solid line; top panel). Bottom panel of Fig.~\ref{fig:hist_modulP} shows the distribution of modulation periods. Although the sample is rather scarce, we may conclude that distribution of pulsation periods of modulated stars follows the overall period distribution well. Most of the modulated stars have pulsation periods in between 1 and 2.5\,d. Except 5 stars with modulation periods above 300\,d, all other have modulation periods below 120\,d; for majority of stars we find $20<P_{\rm mod}<80$\,d.

\begin{figure}
\noindent\includegraphics[width=\columnwidth]{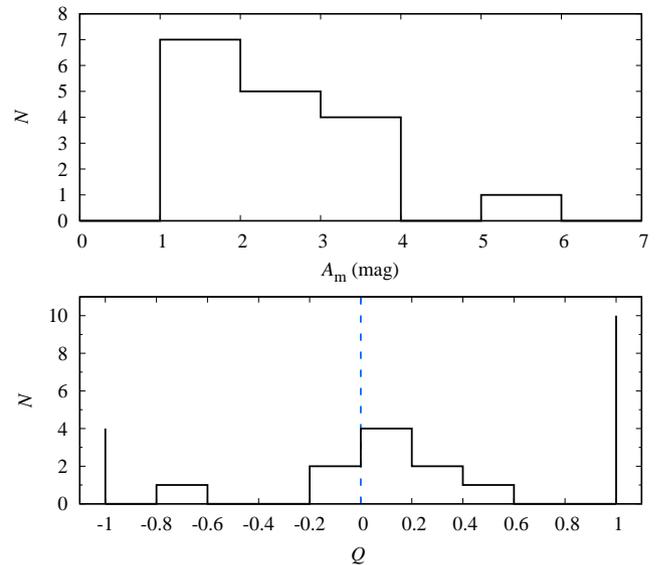}
\caption{Top panel: distribution of amplitudes of mean brightness modulation. Bottom panel: distribution of the asymmetry parameter, $Q$. Vertical bars at $|Q|=1$ indicate the number of stars in which only one side peak was detected at first overtone frequency.}
\label{fig:hist_modulAQ}
\end{figure}

Frequency spectra are scarce, however a glance at Tab.~\ref{tab:mod} reveals some regularities. First, similar to modulated F-mode Cepheids, but in contrast to Blazhko RR~Lyr stars, modulation of mean brightness (a peak at the modulation frequency) is commonly detected (in 17 out of 24 stars). The amplitude of mean brightness modulation, $A_{\rm m}$ (amplitude of the peak at $\fmodul$) is weak, in mmag range. The distribution is plotted in the top panel of Fig.~\ref{fig:hist_modulAQ}. Only in one star it exceeds 5 mmag (\idlmc{2330}). 

The relative modulation amplitude of the radial 1O mode is also typically weak. We define it as $\max(A_+,\,A_-)/\aov$, where  $A_+$ and $A_-$ are amplitudes of side peaks at $\fov+\fmodul$ and $\fov-\fmodul$, respectively. In all but four stars it is below 10 per cent. For the record holder, \idlmc{3985} it is 76\,per cent. Light curve changes in all four stars with relative modulation amplitude exceeding 10\,per cent are visualised with animations attached as Supporting Material. In these stars we were also able to follow the amplitude and phase changes. Panels illustrating modulation in the $A_1-\phi_1$ plane are attached to each animation. In two cases (\idlmc{1705} and \idlmc{3985}) we observe clockwise variation  and in two cases (\idlmc{1474} and \idlmc{4270}) anti-clockwise variation.

In the frequency spectra we observe that doublets (single modulation peak next to $\fov$; 14 stars) are more common than triplets (two side peaks placed symmetrically around $\fov$; 10 stars). Interestingly, for doublets a side peak at $\fov+\fmodul$ is more often detected than the side peak of lower frequency $\fov-\fmodul$. For triplets, we typically find $A_+>A_-$. We illustrate these observations in the bottom panel of Fig.~\ref{fig:hist_modulAQ} in which histogram of the asymmetry parameter, $Q = (A_+-A_-)/(A_++A_-)$ \citep{Alcock2003} is plotted. For doublets $Q$ is equal $+1$ or $-1$. Clearly, positive asymmetry is dominant.

We note, that large-amplitude modulation of very long (unresolved) period is also detected in the analysed sample. An example is \idlmc{4477} which shows clear long-period and irregular modulation on a timescale of 1200-1300\,d, which is well visible already when inspecting photometric data. Nearly two cycles are covered in OGLE-IV data, but in the frequency spectrum the modulation side peaks are unresolved. More data are needed to study this interesting case in more detail.

\begin{table*}
\begin{tabular}{lrrrrrrrr}
   &             &                & \multicolumn{4}{c}{Amplitudes of peaks at: (mag)} & & \\
ID & $\pov$\,(d) & $\pmodul$\,(d) & $\fov$ & $\fmodul$ & $\fov-\fmodul$ & $\fov+\fmodul$ & Additional modulation frequencies. & Remarks \\
\hline
\idlmc{0991} & 3.45244(3)  & 59.4(1)   & 0.0601 &      - & 0.0028 & 0.0032 &  & ap \\ 
\idlmc{1368} & 2.165537(6) & 46.57(9)  & 0.0917 & 0.0012 & 0.0017 & 0.0017 &  & ap, 0.61, nsO, tdp\\ 
\idlmc{1474} & 0.3022899(7)& 34.18(2)  & 0.0539 &      - & 0.0168 & 0.0283 &  &   \\
\idlmc{1626} & 1.803121(2) & 52.03(9)  & 0.1341 & 0.0016 &      - & 0.0016 &  &   \\
\idlmc{1661} & 2.372772(5) & 29.10(2)  & 0.0892 & 0.0035 &      - & 0.0021 &  & ap \\
\idlmc{1685} & 2.667840(5) & 31.85(3)  & 0.0932 & 0.0018 &      - & 0.0012 &  & ap  \\ 
\idlmc{1769} & 2.080187(4) & 33.33(4)  & 0.0809 & 0.0016 &      - & 0.0013 &  & ap \\ 
\idlmc{1815} & 1.927527(3) & 64.1(1)   & 0.0899 & 0.0024 &      - & 0.0017 &  & ap \\ 
\idlmc{1829} & 2.162057(6) & 46.42(7)  & 0.0950 &      - & 0.0021 & 0.0032 & & nsO, ap \\  
\idlmc{2062} & 1.469803(2) & 104.0(3)  & 0.1165 &      - & 0.0026 & 0.0022 & & nsO, ap\\ 
\idlmc{2213} & 2.417636(5) &  80.5(1)  & 0.0870 & 0.0011 &      - & 0.0050 & & nsO \\     
\idlmc{2319} & 2.678569(7) & 42.57(4)  & 0.0900 & 0.0013 &      - & 0.0043 & & nsO \\
\idlmc{2330} & 0.8447634(6) & 111.7(2) & 0.1721 & 0.0052 & 0.0059 & 0.0061 & $2\fmodul$      & nsO \\
\idlmc{2344} & 4.09683(1)  & 36.52(2)  & 0.1146 & 0.0012 & 0.0053 &      - & $2\fov-\fmodul$ & nsO\\
\idlmc{2380} & 2.007854(7) & 66.5(1)   & 0.0736 & 0.0031 &     -  & 0.0018 &  & ap \\
\idlmc{2421} & 2.285858(5) & 79.7(2)   & 0.0978 & 0.0022 &     -  & 0.0035 &  & ap, nsO \\ 
\idlmc{2477} & 1.940064(3) & 483(3)    & 0.1047 &      - & 0.0051 & 0.0020 &  &  \\
\idlmc{2705} & 0.2645433(1)& 330.7(5)  & 0.1024 & 0.0032 & 0.0456 & 0.0092 & $2\fov-\fmodul$, $3\fov-\fmodul$ & ap \\ 
\idlmc{2787} & 3.39640(2)  & 51.56(9)  & 0.0645 & 0.0022 & 0.0014 &      - &  & ap \\ 
\idlmc{2992} & 0.4157542(4)& 473(5)    & 0.1278 &      - & 0.0082 & 0.0104 &  &  \\ 
\idlmc{3226} & 1.467621(3) & 36.82(3)  & 0.0879 & 0.0025 &      - & 0.0050 &  &  weak\\
\idlmc{3985} & 4.54494(6)  & 13.5767(8)& 0.0232 & 0.0029 & 0.0176 &      - & $2\fov-\fmodul$, $3\fov-\fmodul$, $\fov-2\fmodul$ & \\
\idlmc{4270} & 1.430489(3) & 540(2)    & 0.0841 &      - & 0.0086 & 0.0105 & $2\fov-\fmodul$,  $2\fov+\fmodul$ & \\
\idlmc{4361} & 1.805423(5) & 335(3)    & 0.0964 & 0.0031 & 0.0042 &      - &  & \\
\hline
\idsmc{4039} & 1.672049(4)  & 132.2(6)  & 0.1260 & 0.0038  & -  & 0.0042 &  & nsO\\
\idsmc{4241} & 3.29770(3)  & 123.6(1) & 0.0295 & 0.0028 & -  & 0.0022  & $2\fmodul$, $3\fmodul$, $4\fmodul$, $\fov+2\fmodul$, $\fov-2\fmodul$  & \\
\idsmc{4372} & 1.451329(2)  & 943(24) & 0.1451 & - & 0.0035 & 0.0055 & $2\fov+\fmodul$ & ap\\
\hline
\end{tabular} 
\caption{Properties of modulated 1O Cepheids. Consecutive columns contain: star's ID, pulsation period, modulation period, amplitudes of peaks detected at $\fov$, $\fmodul$, $\fov-\fmodul$ and $\fov+\fmodul$, and a list of other modulation peaks detected. Remarks in the last column are: `nsO' -- non-stationary first overtone, `0.61' -- additional periodicity with $\px/\pov\in(0.60,\,0.65)$ was detected, `ap'-- other additional periodicity detected, `tdp' -- time-dependent analysis was conducted.}
\label{tab:mod}
\end{table*}

{\it Remarks on individual stars:}

\idlmc{0991}, $\pov=3.45244(3)$\,d, $\pmodul=59.4(1)$\,d. Modulation properties of this star were discussed in detail in \cite{Kotysz}.

\idlmc{1368}, $\pov=2.165537(6)$\,d, $\pmodul=46.57(9)$\,d. Peaks at $\fmodul$, $\fov-\fmodul$ and $\fov+\fmodul$ were detected with $\sn=4.0$, $4.6$ and $4.6$, respectively, but only after time-dependent prewhitening was applied due to non-stationary nature of first overtone mode. In the same star, we detect signals from classes discussed in Sect.~\ref{ssec:61} and in Sect.~\ref{ssec:unclassified}.

\idlmc{1474}, $\pov=0.3022899(7)$\,d, $\pmodul=34.18(2)$\,d. A triplet at radial mode frequency is clear, with $\sn=10.7$ and $7.5$ for peaks at $\fov-\fmodul$ and $\fov+\fmodul$, respectively, but this triplet represents the only significant power in the frequency spectrum. No harmonics of 1O are detected. The star holds one of the shortest periods in the whole sample and large relative modulation amplitude of 53\,per cent. Light curve changes along the modulation cycle are illustrated with an animation available as Supporting Information.


\idlmc{1661}, $\pov=2.372772(5)$\,d, $\pmodul=29.10(2)$\,d. Peaks at $\fmodul$ and at $\fov+\fmodul$ detected with $\sn=9.5$ and $5.9$, respectively -- see Fig.~\ref{fig:fspmodul}. Clear mean brightness modulation. Two additional significant signals detected in the frequency spectrum; the one close to $\fov$ may indicate secondary modulation with $\pmodul\approx 743$\,d.




\idlmc{1829}, $\pov=2.162057(6)$, $\pmodul=46.42(7)$. Non-stationary first overtone with strong residual of $\sn=9.2$ after prewhitening. Peaks at $\fov-\fmodul$ and at $\fov+\fmodul$ detected with $\sn=5.9$ and $5.4$, respectively -- see Fig.~\ref{fig:fspmodul}. The signal at $\fov-\fmodul$ is non-coherent. Additional signals detected as well.




\idlmc{2330}, $\pov=0.8447634(6)$\,d, $\pmodul=111.7(2)$\,d. Triplet components at $\fov+\fmodul$ and $\fov-\fmodul$ are detected with $\sn$ of $6.4$ and $6.1$ respectively. Then, signals at the modulation frequency and its harmonic are detected with $\sn=5.7$ and $4.3$, respectively -- see Fig.~\ref{fig:fspmodul}. First overtone is non-stationary. A weak ($\sn=3.2$) signature of quintuplet component ($\fov+2\fmodul$) is visible in the spectrum.

\idlmc{2344}, $\pov=4.09683(1)$\,d, $\pmodul=36.52(2)$\,d. A peak at $\fov-\fmodul$ is prominent with $\sn=11.7$. After removing low frequency power excess with two low frequency ($\nu<0.01\cd$) sines weak modulation peaks at $2\fov-\fmodul$ ($\sn=3.9$) and at $\fmodul$ ($\sn=3.4$) are detected. While for these peaks $\sn<4$ the simultaneous presence of three modulation-related peaks prompts us to classify the star as a modulation candidate.




\idlmc{2705}, $\pov=0.2645433(1)$\,d, $\pmodul=330.7(5)$\,d. A very clear modulation of large amplitude with rich frequency spectrum. The strongest modulation signals at $\fov-\fmodul$, $\fov+\fmodul$ and $2\fov-\fmodul$ are detected with $\sn=17.2$, $6.8$ and $5.5$, respectively. Light curve changes along the modulation cycle are illustrated with an animation available as Supporting Information.




\idlmc{3985}, $\pov=4.54494(6)$\,d, $\pmodul=13.5767(8)$\,d. Extreme case of large amplitude modulation, with peculiar light curve changes during the modulation cycle. The strongest modulation peaks are detected at $\fov-\fmodul$, $2\fov-\fmodul$, $\fmodul$ and $3\fov-\fmodul$ with $\sn=22.8$, $10.0$, $9.7$, and $8.3$, respectively. Light curve changes along the modulation cycle are illustrated with an animation available as Supporting Information.

\idlmc{4270}, $\pov=1.430489(3)$\,d, $\pmodul=540(2)$\,d. A very clear long period amplitude modulation. Triplet components at $\fov-\fmodul$ and at $\fov+\fmodul$ detected with $\sn=13.1$ and $14.0$, respectively. Light curve changes along the modulation cycle are illustrated with an animation available as Supporting Information.

\idlmc{4361}, $\pov=1.805423(5)$\,d, $\pmodul=335(3)$\,d. Since 1-yr alias of a signal detected at $\fov-\fmodul$ is unresolved with $\fov$ the case is a bit ambiguous. Peaks at $\fov-\fmodul$ and $\fmodul$ are firmly detected with $\sn=5.4$ and $4.6$, respectively.


\idsmc{4241}, $\pov=3.29770(3)$\,d, $\pmodul=123.6(1)$\,d. A very peculiar star. No harmonics of first overtone are detected in the frequency spectrum. After prewhitening with $\fov$ a comb of strong peaks is visible at low frequencies, at $\fmodul$, $2\fmodul$ and $3\fmodul$ with $\sn=6.1$, $6.4$ and $5.1$, respectively, and at $\fov+\fmodul$ ($\sn=4.8$) -- see Fig.~\ref{fig:fspmodul}. In consecutive prewhitening steps we detect additional weak modulation peaks (all with $3.5<\sn<4$) at $4\fmodul$, $\fov+2\fmodul$ and $\fov-2\fmodul$. All signals are present when data are transformed into fluxes, which proves both pulsation and modulation frequencies originate from the same star. The star appears to have a very peculiar modulation with strongly nonlinear mean brightness variation. For ground-based data, detection of $2\fmodul$ in Blazhko RRab stars is very scarce; for \idsmc{4241} we detect harmonics up to $4\fmodul$. Light curve changes along the modulation cycle are illustrated with an animation available as Supporting Information.

\idsmc{4372}, $\pov=1.451329(2)$\,d, $\pmodul=943(24)$\,d. Star shows weakly resolved, long-period modulation. Two modulation peaks at $\fov-\fmodul$ and $\fov+\fmodul$ are detected with $\sn=6.8$ and $7.6$, respectively. Then, after prewhitening we detect one more modulation component at $2\fov+\fmodul$ ($\sn=4.3$) and additional unclassified periodicity.

\section{Discussion}\label{sec:discussion}

\subsection{Origin of the observed additional periodicities}

In Sect.~\ref{sec:results} we have divided the detected additional periodicities into a few classes based on their period ratios with respect to radial first overtone period, nature of the additional periodicity or the appearance of additional structures in the frequency spectra. For these classes, origin of the additional variability likely differs. Before discussing possible mechanisms behind excitation of additional variability, we first note that some of the detected signals may originate due to contamination, ie., are not intrinsic to Cepheid, but originate from other variable star in its direct neighbourhood. Signals we detect form regular patterns in the Petersen diagram, or are characterised by distinct properties, which is not expected due to contamination. We may thus conclude that majority of signals we detect are intrinsic to the studied Cepheids. Possible contamination, while may add some `noise', eg., to the appearance of the Petersen diagrams, is not expected to alter statistical properties of the classes we have identified.

\subsubsection{Double-periodic stars with $\px/\pov$ in between 0.60 and 0.65}
\label{ssec:discussion61}

\cite{wd16} proposed that the three sequences observed in the Petersen diagram (Fig.~\ref{fig:pet61}) are due to harmonics of non-radial modes of moderate degrees, $\ell=7$ (top sequence), 8 (middle sequence) and 9 (bottom sequence). Harmonics should be more easy to detect due to strongly reduced geometric cancellation. Non-radial modes can also be detected; assuming intrinsic mode amplitude is the same, modes of $\ell=8$ should reach the highest amplitudes followed by modes of $\ell=7$ and $\ell=9$. Non-radial modes are also expected to have more complex appearance in the frequency spectrum due to nonlinear mode interactions within multiplets, ie., between modes of different azimuthal orders. This is what we observe (Fig.~\ref{fig:pet61}, Tab.~\ref{tab:seqprop}). In the nomenclature of Sect.~\ref{ssec:61} non-radial modes correspond to sub-harmonics, with amplitudes ($\ash$) determined by the highest peak within wide power band that is commonly observed in the frequency spectrum. Sub-harmonics are most often detected for the middle sequence, then for the top sequence and they are scarce for signals forming the bottom sequence. Differences are clear and significant for the SMC. We note that for LMC, while sub-harmonics are still more frequent for the middle sequence, the incidence rate is comparable to that of the top sequence. These observations support \cite{wd16} model. The fact that we observe so many non-radial modes directly is puzzling however, and challenging to explain, as their observed amplitudes should be strongly reduced due to cancellation. In fact in the LMC stars for which the amplitude of non-radial mode is larger than that of the harmonic, $\ash>\ax$, dominate (Fig.~\ref{fig:61shamps}). In the next section we even suggest that many of signals with period ratios in between $0.75$ and $0.90$ may also be non-radial modes of the same degrees, for which harmonics are not detected at all. This, together with significant dispersion of period ratios in the Petersen diagram in the LMC may be challenging to reconcile within \cite{wd16} model. 

The tentatively identified 4th sequence would, in light of \cite{wd16} model, correspond to $\ell=6$ non-radial modes. It includes 2 stars only, however, and detailed calculations for $\ell=6$ modes were not conducted so far.

We note that even for RRc stars, {\it TESS} mission reveals double-periodic stars with period ratios significantly below 0.61, possibly indicating the excitation of $\ell=10$ modes \citep[see fig.~13 in][]{MolnarTESS}.

\subsubsection{Non-radial $\ell=7-9$ modes without harmonic?}
\label{ssec:789noharm}

In Sect.~\ref{ssec:unclassified} we have identified a broad peak ($0.75<\pslr<0.90$) in the distribution of period ratios for non-coherent signals, with $\px>\pov$ (for LMC, see Fig.~\ref{fig:histadd}). A group of double-periodic stars of similar properties was detected by \cite{sa18b} (called 1.25-mode in their paper). These authors discussed the hypothesis that these signals may correspond to non-radial modes of degrees 7, 8 and 9. We put the same hypothesis and suggest that these stars are directly coupled to stars with $\px/\pov$ in between $0.60$ and $0.65$. A glance at Fig.~\ref{fig:pet61} shows that such period ratios are expected for non-radial modes of degrees 7, 8 and 9, see preceding Section. We also note that these non-radial modes most frequently appear as non-coherent signals. In Fig.~\ref{fig:pet61} we have defined two polygons (dashed lines), separately for the SMC and the LMC, that enclose majority of stars with non-radial modes of degrees 7, 8 and 9. The same polygons are also plotted in Fig.~\ref{fig:petadd} and we can see that the broad peak in the distribution of period ratios in Fig.~\ref{fig:histadd} is mostly due to stars within these polygons. We note that coherent signals also fall within these polygons and those in Fig.~\ref{fig:pet61} do correspond to non-radial modes of degrees 7, 8 and 9.

We speculate that majority of signals that fall within the polygons in Fig.~\ref{fig:petadd} are in fact non-radial modes of degrees 7, 8 and 9, for which harmonic was not detected in the frequency spectrum. These may be both non-coherent signals (which is more frequent) and coherent ones. Among stars with $\px/\pov$ in between 0.60 and 0.65 amplitudes of non-radial modes (denoted by $\ash$ in Sect.~\ref{ssec:61}) are often larger than amplitudes of their harmonics (denoted by $\ax$ in Sect.~\ref{ssec:61}). In fact, in the LMC $\ash>\ax$ is more common -- see Fig.~\ref{fig:61shamps}. Hence, it is likely, that we may just see non-radial mode directly, but not its harmonic, eg., due to large noise level.

\begin{figure}
\includegraphics[width=\columnwidth]{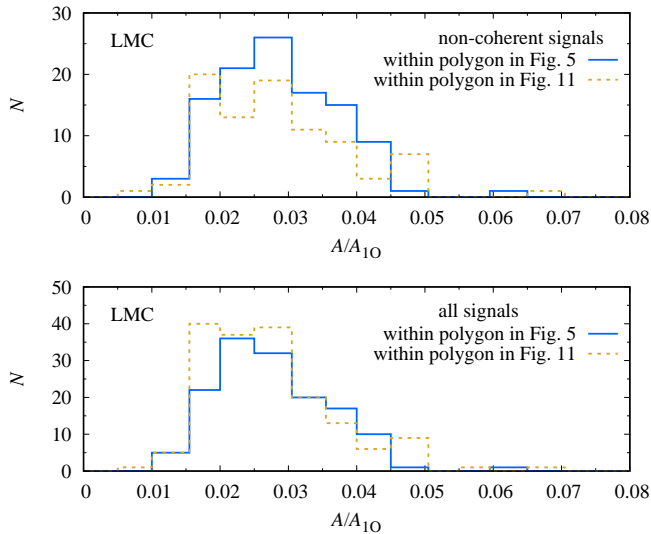}
\caption{Distribution of relative amplitudes for signals within polygons plotted in Fig.~\ref{fig:pet61} (solid line) and in Fig.~\ref{fig:petadd} (dashed line). Top panel shows distribution for non-coherent signals, while bottom shows distribution for all signals.}
\label{fig:polygon}
\end{figure}

In Fig.~\ref{fig:polygon} we plot the distribution of relative amplitudes for signals with $\px>\pov$ that fall within the polygons plotted in the right panels of Figs~\ref{fig:pet61} and \ref{fig:petadd}. We use data on more numerous LMC sample only. In the top panel we plot the distribution for non-coherent signals, while in the bottom panel we plot the distribution for all signals. With solid line we show the distribution for stars with $\px/\pov$ in between 0.60 and 0.65 (stars within polygon in Fig.~\ref{fig:pet61}), while dotted line shows the distribution for stars for which harmonics were not detected (signals within polygon in Fig.~\ref{fig:petadd}). Despite some differences, the distributions are similar and cover the same range, which also applies to the broad maxima. It strengthens the hypothesis that all these signals arise due to the same phenomenon.

\subsubsection{Close non-radial modes?}
\label{ssec:cnm}

Another group identified in Sect.~\ref{ssec:unclassified} constitutes of stars with additional coherent variability, majority of which have large period ratios ($\pslr>0.95$), ie., their frequencies are close to the radial mode frequency. For such signals, their relative amplitude is also typically larger (Fig.~\ref{fig:ampadd}). These signals cannot be explained as due to radial modes. Two explanations are possible. Additional variability may arise due to periodic modulation of a radial mode. Then, we expect to see triplets (multiplets in general) centred on radial mode frequency, or side peaks located on one side of the radial mode frequency and its harmonics (so called doublets; see Sect.~\ref{ssec:modulation}). In the discussed stars we see only a single peak. One may speculate that other modulation peaks have too small amplitudes, below the detection limit. This explanation is not likely however. We note that the same speculation motivated \cite{Kotysz} to revisit the stars sharing the above properties, first analysed by \cite{mk09} with OGLE-II data, to look for missing modulation components. They used more extended OGLE-III/IV data with significantly lower noise levels and found a signature of weak modulation in only one out of 42 stars. The class we address also outnumbers the modulated stars we have detected. If these single peaks were due to modulation, its properties would have to be very peculiar, with extremely asymmetric side peaks. 

The more likely explanation seems excitation of low-degree non-radial modes with frequencies close to the radial mode frequency. Such scenario however is difficult to reconcile with pulsation theory. While low-order non-radial modes are expected to be excited in RR~Lyr stars \citep{VanHoolst,DziembowskiCassisi}, for classical Cepheids non-radial modes with $\ell<4$ are expected to be damped \citep{wd77,Osaki77,MuletMarquis}.

\subsubsection{Stars with period ratios $\pov/\px$ around 0.684}

This group is primarily known in RR~Lyr stars. In classical Cepheids it was first reported by \cite{sa18b} (called 1.46-mode in their paper). We confirm its existence. In the Petersen diagram it is not as distinct as is the case for RR~Lyr stars, still the distribution of period ratio leaves little doubt that also in 1O Cepheids the class exists, showing a bit smaller median period ratio, $0.684$ instead of $0.686$ in RRc stars. The basic difficulty in explaining this class is long period of the additional periodicity -- longer than period of the radial fundamental mode expected for a corresponding first overtone period. Except a few comments on RRc stars \citep{wd16} no model was proposed to explain this class. A model should explain existence of this class in both RRc and 1O Cepheids, narrow distribution of period ratios and coherent nature of the additional variability.

\subsubsection{Stars with periodic modulation of pulsation}

Periodic modulation of pulsation is a well know feature of RR~Lyr stars and is called the Blazhko effect \citep{Blazhko}. It is common among fundamental mode pulsators \citep[incidence rates on order of 50\,per cent, for a review see][]{GezaBlazhkoIR}, less common in first overtone ones \citep[incidence rates of a few per cent, see eg.,][]{NetzelBlazhkoRRc} and detected also in double-mode (RRd) pulsators \citep[eg.,][]{JurcsikM3RRd,SmolecBlazhkoRRd}. Detections of periodic modulation of pulsation in classical Cepheids are scarce, however new discoveries have been appearing regularly in recent years. These include modulations of radial modes in double-mode Cepheids \citep{mk09, Rathour2021} and single-mode Cepheids \citep{s17, Rathour2021}. For single-mode overtone Cepheids, mostly long-period, large amplitude modulations, well visible already in the light curve were reported \citep[eg.,][]{MolnarSzabados,o4_clouds_multimode}. Here we reported several modulated 1O Cepheids based on the appearance of their frequency spectra. The existence of periodic modulation of pulsation in Cepheids is now firmly established. We refrain from calling the effect `Blazhko'. Certainly it is much more scarce as compared to RR~Lyr stars. In majority of cases it is a low-amplitude effect. The mechanism behind might be different.

We still lack satisfactory explanation behind Blazhko modulation in RR~Lyr stars, although the half-integer resonance model seems the most promising one -- see \cite{KollathRev} for a review. In \cite{s17} we speculated that a resonance of a different type (2:1 between radial fundamental and radial second overtone) may be crucial in bringing modulation in F-mode Cepheids, which was motivated with the highest incidence rate of modulated stars close to the resonance center. We don't see an obvious resonance, or any other mechanism that might work for 1O Cepheids.

\subsection{Properties of 1O Cepheids in the LMC and SMC}
\label{ssec:propSLMC}

A glimpse at the Petersen diagram in Fig.~\ref{fig:petall} reveals a significant difference between the two Magellanic Clouds: additional variability beyond radial modes is detected much more frequently in the LMC. For all discussed classes of pulsations, incidence rates are higher in the LMC. The difference is intrinsic and is not caused by the higher detection limit in the SMC, as we have demonstrated in Sect.~\ref{ssec:unclassified}. Thus, higher metallicity environment seems to favour excitation of additional variabilities beyond radial modes.

Except different incidence rates, different properties within the discussed classes were also reported and are most prominent in double-periodic stars with $\px/\pov$ in between 0.60 and 0.65. We observe the following. (i) Double-periodic stars in the LMC have on average longer period. This trend with increasing metallicity was already noted by \cite{Rathour2021} who also analysed classical Cepheids in the Galactic fields. (ii) The appearance of the Petersen diagram is much more `noisy' for LMC stars. The three sequences, although still well visible, are not as clearly cut as in the SMC. There are several stars in between the sequences. (iii) For stars in which signals at $\fx$ and centred at $1/2\fx$ were detected, distribution of corresponding amplitudes, $\ax$ and $\ash$, respectively, is different. While in the SMC $\ax>\ash$ is more frequent, in the LMC $\ax<\ash$ is more frequent.

Another striking difference is the appearance of the modulations in the two systems. First, the incidence rate of modulated stars is very low in the SMC: we have detected only 3 modulated stars there, while 24 were detected in the LMC. By applying the SMC's detection limit, to LMC stars, we still should detect 8 modulated stars (ie., at least two modulation side peaks would be detected). 

While the statistics is low, we note that stars in the SMC have, on average, longer modulation periods. 

Any model aimed at explaining the different classes of pulsations we have described, should also explain the metallicity/environmental dependencies.

\section{Conclusions}

We have analysed OGLE-IV $I$-band photometry for first overtone classical Cepheids in the Magellanic Clouds. Our most important findings are the following.

(i) We have identified new candidates for radial double-mode pulsation, including 46 candidates for F+1O pulsation, 21 candidates for 1O+2O pulsation and 4 candidates for 1O+3O pulsation.

(ii) In 225 stars in the SMC and 291 stars in the LMC we have detected additional signals that form period ratios with respect to first overtone $\px/\pov$ in the 0.60 -- 0.65 range. In the Petersen diagram these stars form three sequences -- clearly cut in the case of SMC and more diffuse in the LMC. Quantitative method to divide stars into sequences was proposed. For a significant fraction of these stars, signals centred at sub-harmonic frequencies are detected as well, which according to model proposed by \cite{wd16} are non-radial modes of moderate degrees, 7, 8 and 9. Distribution of these signals supports the Dziembowski's model. In the light of this model, the distribution of amplitudes of the detected signals is puzzling. Amplitudes of non-radial modes should be strongly reduced due to geometric cancellation, and their harmonics should be much more easy to detect. In the LMC we find that amplitudes of non-radial modes are typically larger than amplitudes of their harmonics.

(iii) Two stars were tentatively classified as belonging to the fourth sequence with highest $\px/\pov$ period ratios. Extending the \cite{wd16} model, the fourth sequence would arise because of non-radial $\ell=6$ modes.

(iv) We have identified 28 stars with additional variability of period longer than period of the radial fundamental mode expected for a given 1O period. Period ratios tightly cluster around $\pov/P=0.684$. Additional signals are coherent; typical amplitude is 3\,per cent of the radial 1O amplitude. In Cepheids, this group was first discovered by \cite{sa18b} and is analogue of double-periodic pulsation known in first overtone RR~Lyr stars.

(v) After clean-up of the sample from signals corresponding to the groups discussed above, hundreds of signals are still present across the Petersen diagram. The following regularities were spotted. For majority of signals frequencies are smaller than the frequency of the radial first overtone. Distribution of period ratios peaks for $\pslr>0.95$, then we observe a plateau down to $\pslr=0.75$ and decline further on. Coherent signals (of constant amplitude and phase) constitute the vast majority of signals close to the radial mode frequency ($\pslr>0.95$). These signals have, on average, larger relative amplitudes. The plateau is built up by both coherent and non-coherent signals. For non-coherent signals we observe a broad maximum for period ratios in between 0.75 and 0.9. We suggest that coherent signals with close to unity period ratios may be low-order non-radial modes. Signals with lower period ratios, in particular non-coherent ones, may represent non-radial modes of degrees 7, 8 and 9, as first suggested by \cite{sa18b}. Both explanations pose a challenge for theory, as low order non-radial modes are linearly damped, and for moderate-degree modes amplitudes should be strongly reduced by geometric cancellation.

(vi) In many stars various forms of multiperiodic pulsation are present simultaneously. These include eg., radial double-mode stars with additional signals with $\px/\pov$ in the $0.60-0.65$ range or clustered around $\pov/\px=0.684$. The latter two forms are also present simultaneously in two stars. All these stars are promising targets for detailed asteroseismic investigation.

(vii) In the frequency spectra of 27 stars (24 in the LMC, 3 in the SMC) we detect periodic modulation of pulsation of radial first overtone. Modulation periods are in between $\approx 13$\,d and $\approx950$\,d. Relative modulation amplitudes may be as high as 76\,per cent. Periodic modulation of pulsation is now firmly established in nearly all classes of single and double-mode pulsation in Cepheids. Still phenomenon is much more scarce than in RR Lyrae stars, in which it is known as the Blazhko effect.

(viii) Incidence rates for all discussed phenomena are larger in the more metal rich LMC. We have demonstrated that it does not result from higher detection limit in the SMC, but is intrinsic property of 1O Cepheids. Higher metallicity environment favours excitation of additional variabilities.

\section*{Acknowledgements}
This research was supported by the Polish National Science Centre, SONATA BIS grant, 2018/30/E/ST9/00598 and OPUS grant, DEC-2015/17/B/ST9/03421. RS acknowledges fruitful discussion with Pawe\l{} Moskalik.

\section*{Data Availability}
OGLE data analysed in this paper are publicly available through OGLE Collection of Variable Stars. Basic properties of the discussed stars are available through Tables published online along with the paper. Full frequency solutions for analysed stars are available on request.



\bibliographystyle{mnras}
\bibliography{paper_clean} 


\section*{Supporting Information}
Supplementary data are available at {\it MNRAS} online.

{\bf Animations.} For five stars described in Section~\ref{ssec:modulation} animations of light curve changes across the modulation cycle are available.

{\bf Table~A1}. Properties of double-periodic stars with $\px/\pov$ in between 0.60 and 0.65.

{\bf Table~A2}. Additional properties of double-periodic stars with $\px/\pov$ in between 0.60 and 0.65 for which sub-harmonics were detected.

{\bf Table~A3}. Properties of 1O Cepheids with additional variability detected.




\bsp	
\label{lastpage}
\end{document}